\documentclass[aps,showpacs,twocolumn,superscriptaddress,floatfix]{revtex4}

\usepackage{graphicx}

\def\nbZ{{\mathchoice {\hbox{$\sf\textstyle Z\kern-0.4em Z$}}
{\hbox{$\sf\textstyle Z\kern-0.4em Z$}} {\hbox{$\sf\scriptstyle
Z\kern-0.3em Z$}}  {\hbox{$\sf\scriptscriptstyle Z\kern-0.2em Z$}}}}

\begin{document}

%
%
\title{Discrete non-Abelian gauge theories in Josephson-junction arrays and quantum computation}
%
%
%
\author{Benoit Dou\c{c}ot}
\affiliation{Laboratoire de Physique Th\'{e}orique et Hautes
\'Energies, CNRS UMR 7589,
Universit\'{e}s Paris 6 et 7, 4, Place Jussieu, 75252 Paris Cedex 05, France}

\author{Lev B. Ioffe}
\affiliation{Center for Materials Theory,
Department of Physics and Astronomy, Rutgers University,
136 Frelinghuysen Road, Piscataway New Jersey 08854, USA}

\author{Julien Vidal}
\affiliation{Groupe de Physique des Solides, CNRS UMR 7588,Campus Boucicaut, 
140 rue de Lourmel, 75015  Paris, France}
%
%

\begin{abstract}

We discuss real-space lattice models equivalent to gauge theories
with a discrete non-Abelian
gauge group. We construct the  Hamiltonian formalism which is
appropriate for their
solid-state physics implementation and outline their basic properties.
The unusual physics of these systems is due to local constraints on
the degrees of freedom
which are variables localized on the links of the lattice.
We discuss two types of constraints that become
equivalent after a duality transformation for Abelian theories but
are qualitatively
different for non-Abelian ones.  We emphasize highly nontrivial
topological properties of the
excitations  (fluxons and charges) in these non-Abelian discrete
lattice gauge theories. 
We show that an implementation of these models may provide one with the
realization of an ideal quantum computer, that is the computer that is
protected from the noise and allows a full set of precise manipulations
required for quantum computations.  
We suggest a few designs of the Josephson-junction arrays that provide the 
experimental implementations of these models and discuss the physical 
restrictions on the parameters of their junctions.

\end{abstract}

%
%
%


\maketitle
%
%
%

\section{Introduction}

%
%
%
The ideal quantum computer \cite{Ekert1996,Steane1998}, if it exists,
is a very interesting
object from a physical point of view. It is a system which may be
prepared in any
superposition of $2^K$ (with $K\sim 10^{4}-10^{6}$) basic states during a very
long time without decoherence. To give an order of magnitude,
the decoherence should be much smaller than $10^{-6}$ after a
characteristic time scale
of one operation \cite{Preskill1998}.
Furthermore, it should allow a set of controlled unitary operations
with an accuracy of
$10^{-6}$ or better. It is natural to think that in order to satisfy
these stringent criteria
one needs to have a physical system in which decoherence and errors
generated by manipulations
are exponentially small with respect to the system size  or to any
other control parameter.
We shall refer to such properties as ``ideal" in  the following. The
requirement of an ideal
decoherence can be achieved in physical systems with topological
order parameters
\cite{Wen1990,Wen1991,Read1991} which can be implemented in Josephson-junction
arrays \cite{Ioffe2002,Doucot2003}. However, the physical arrays
proposed so far do not
allow the ideal implementation of the full set of operations
\cite{Ekert1996,Steane1998}
required for quantum computations. In fact, these arrays allow to
perform ideally only the simplest operations: a flip of an individual
bit and the change of the wave function phase by $\pm 1$.
It is crucial that they do not allow precise operations that mix the classical
states, e.g., the Hadamard rotation $U_H$ that transforms the states
$|0\rangle$, $|1\rangle$ into their mixtures
$(|1\rangle\pm|0\rangle)/\sqrt{2}$
requires an application of the correct external perturbation $V$ for a
given time $t$: $U_H=\exp(-iVt)$ and thus is subjected to errors in both the
values of $t$ and $V$.

Generally, an ideal operation involves an adiabatic change of some parameters that results in a generalized Berry's phase.
Note that all alternative ways to operate the qubit are similar to a spin rotation in an external field. Such operations are not adiabatic
and are controlled by the duration of the field pulse which can never be made precise. In fact, it is very difficult to imagine how one
can possibly achieve the accuracy of the transformation that is needed for quantum computation with this technique. In contrast,
adiabatic transformations can be made, in principle, very precise. The  main difficulty with them is to find a system in which the
group formed by these transformation is big enough for a quantum computation.
The arrays discussed above are implementing the symmetry group $\nbZ_2$ and thus it is not
surprising that for them adiabatic changes of the external parameters allow only a very limited set
of operations.
One expects, following the suggestion by Kitaev \cite{Kitaev1997}, that one would get much larger sets of operations in systems
allowing more complicated local symmetry transformations, in particular, in the case of discrete non-Abelian groups \cite{Propitius1999}.

The purpose of this paper
is to consider lattice gauge theories built on more
complicated groups than $\nbZ_2$ and to identify some that allow
nontrivial adiabatic transformations. After having described the
most promising ones, we discuss how they can be implemented, at least,
in principle, in
Josephson-junction arrays.
All solid-state physics implementations of lattice gauge theories
involve a lattice
in real space but imply a continuous time. The natural framework to
describe such
models and their experimental realizations is the Hamiltonian formalism.
This calls for an explicitly time-independent dynamics which can be
achieved by choosing
the scalar part of the gauge field  $A_0=0$. In this case, the vector
part $A_{ij}$ describes
the quantum state of the physical links. However, the Hamiltonian
thus constructed still
commutes with the large group of local time-independent gauge transformations.
It is then convenient to use a description which keeps track only of
the underlying  physical
degrees of freedom. This can be done by restricting the Hilbert space
to the space of gauge-invariant states.
Note that several studies have already pointed out the existence of close
connections between Josephson-junction arrays, lattice gauge theories and generalized
statistics for elementary excitations. For instance, incompressible phases analogous
to the fractional quantum Hall fluids (FQHF) of two-dimensional electronic systems
have been predicted and studied in such arrays. One of them may be viewed as a FQHF
of charges \cite{Odintsov} induced by an external magnetic field, whereas the other
involves a FQHF of vortices \cite{Choi,Stern} in the presence of a uniform fractional
background charge on the superconducting islands. An appealing synthesis of these possible
phases in the language of lattice gauge theories has been developed in Ref.
\cite{Diamantini}. As we shall show in detail, the lattice gauge models discussed here have
the main advantage over the previous ones that effective interactions between elementary
excitations are purely of statistical origin, and do not, for instance, depend on their
precise locations. This is, of course, highly desirable in the context of quantum
computation where the goal is to implement, with high accuracy, a set of universal
unitary transformations.

This paper is organized as follows. In Sec. \ref{CGT}, we review
basic notions of continuous gauge groups and introduce basic notations. We
emphasize the connection between the gauge-invariance constraint
imposed on the physical states and the Gauss law constraint. Special
features arising for non-Abelian groups are discussed.
Section \ref{DAGT} focusses on the Abelian cyclic $\nbZ_n$ group.
In this case, we show that imposing a local constraint on the product
of group elements along the links of elementary plaquettes (kinematical
Gauss law) is equivalent, after a duality transformation, to the requirement
of local gauge invariance. We briefly discuss implementations of these
two versions of $\nbZ_n$ models in Josephson-junction arrays and the
existence of a topologically protected ground state when arrays exhibit
macroscopic holes. Section \ref{DNAGT} addresses the interesting case
of theories associated with a finite non-Abelian group. In this situation,
the usual local gauge-invariance constraint and the kinematical Gauss
law constraint have to be treated separately, since no exact duality
transformation is available to connect them. We show that both models
present gapped low-energy excitations called fluxons which may be
trapped in macroscopic holes of the arrays. We also discuss local
violations of the constraints called pseudocharges, and show the existence
of a non-Abelian Aharonov-Bohm effect as a pseudocharge winds around a fluxon
(and conversely). Section \ref{Implementations} is dedicated to
implementations of these two models, for the simplest non-Abelian
groups such as the dihedral groups $D_3$ and $D_4$. These proposals are
based on the frustrated Josephson-junction arrays, where some basic
elements have a $2\pi/n$ periodic potential energy as a function
of the phase difference across them. The computations of the  various energy
scales driving the physical properties of these arrays, are presented in four
appendixes.
%
%
%

\section{Continuous gauge theories}
\label{CGT}

%
%
%
Let us start by a brief review of the properties of lattice gauge theories with
continuous groups using the language that will be convenient for the
discussion of discrete groups. Therefore, we consider a Hamiltonian $H$ and a
Hilbert space that are
invariant under the local gauge transformations built from
a continuous group $G$.

More precisely, to any set of group elements $\{g_i\}$'s labeled by
lattice sites, we
associate the transformation $U_{\{g\}}$ such that
$|\Psi \rangle \rightarrow U_{\{g\}} |\Psi  \rangle=|\Psi \rangle$,
and $[H,U_{\{g\}}]=0$.
For a given lattice, the gauge field $A$ is attached to the links and
takes its values in  the group $G$. Local gauge transformations are
then given by
%
%
\begin{equation}
A_{ij} \rightarrow {g}_i A_{ij} {g}_j^{-1} \equiv A'_{ij}.
\end{equation}
%
In the familiar example of compact quantum electrodynamics (QED)
$A_{ij}=\exp(i a_{ij})$,
$g_i=\exp(i \phi_i)$ and gauge transformation in the variables ($a_{ij}$,
$\phi_i$) acquires a well-known form $a_{ij} \rightarrow a_{ij} +
\phi_i - \phi_j$.
Identifying quantum states $|\Psi \rangle$ with functions $\Psi$ of
classical field configurations
$\{A_{ij} \}$, the condition of gauge invariance can be explicitly written as
%
%
\begin{equation}
\Psi \{A_{ij} \} = \Psi \{ A'_{ij} \}
\mbox{,}
\label{GaugeInv}
\end{equation}
%
%
for any choice of $g_i$'s. In particular, we may choose
transformations localized on a
single site $i$, so that $g_j=e$ if $j \neq i$ and $g_i=g$,
where $e$ denotes the
neutral element in the group $G$.  Furthermore, since $G$ is a
continuous group,
it is possible to consider infinitesimal $g$'s. The constraint of local gauge
invariance (\ref{GaugeInv}) can then be expressed as
%
%
\begin{equation}
\sum_{j(i)} \frac{\partial}{\partial a_{ij}^\alpha} |\Psi \rangle= 0
\mbox{,}
\label{Gauss}
\end{equation}
%
%
where $a_{ij}^\alpha$ is the generator of the Lie group corresponding
to the direction $\alpha$ in
the vicinity of the neutral element of $G$. In the following, the
notation $j(i)$ stands for all
sites $j$ connected to $i$. The operator conjugated to
$a_{ij}^\alpha$ is  analogous to an
electric field 
$e_{ij}^\alpha = -i \partial / \partial a_{ij}^\alpha$. 
In the  basis where the electrical field
$e^\alpha$ is diagonal, Eq.
(\ref{Gauss})  becomes a lattice equivalent of
the usual Gauss law $\nabla e^\alpha = 0$ which also reads
%
%
\begin{equation}
\prod_{j(i)} E_{ij}^\alpha |\Psi \rangle = |\Psi \rangle
\mbox{,}
\label{DiscreteGauss}
\end{equation}
%
%
where we have set $E_{ij}^\alpha=\exp(ie_{ij}^\alpha t)$, $t$ being
an arbitrary
real parameter. Notice that each site of the lattice corresponds to
plaquette of the dual
lattice, and the product over the links emanating from a site becomes
a product over the links
encircling the corresponding plaquette of the dual lattice. This
allows for a natural realization
of conditions  (\ref{GaugeInv}) and (\ref{DiscreteGauss}) in physical systems.

At this stage, it is very important to point out a deep difference
between Abelian and
non-Abelian groups. In the former case, we may diagonalize simultaneously all
the generators $E_{ij}^\alpha$ since they mutually commute, which is impossible
in the non-Abelian case.
Because of this, we shall explore two different routes in this work.
A natural possibility is to regard Eq. (\ref{GaugeInv}) as the basic
equation, and as
we shall show, it is easy to use this construction for {\em discrete}
non-Abelian groups. A
second approach is to take Eq. (\ref{DiscreteGauss}) as the most
fundamental. In  this case,
we assume that the physical states are also generated by classical
configurations of group elements $E_{ij}^\alpha$ attached to the
links, and we require
these classical configurations to satisfy the  \textit{kinematical} constraint
%
%
\begin{equation}
\prod_{j(i)} E_{ij}^\alpha = e
\mbox{.}
\label{ModifiedGauss}
\end{equation}
%
%
Note that in this new scheme, and if $G$ is non-Abelian, we can no
longer interpret
this modified Gauss law as a consequence of a local gauge symmetry.
So, these two approaches
yield essentially different theories, except in the Abelian case where
their equivalence can be precisely established as explained in the
following section.
%
%
%

\section{Discrete Abelian gauge theories}
\label{DAGT}
%
%
%
%
Let us start with the usual Abelian case and consider  more specifically
a simple example where the discrete gauge group is the cyclic group
$\nbZ_n$, which has also
the major advantage that it can be easily realized in a physical
system. Indeed, this can
be implemented with Josephson-junction elements for which the
Josephson Hamiltonian is
$2\pi/ n$ periodic in the phase  difference $(\phi_a-\phi_b)$ where
$\phi_{a}$ denotes the
phase of the superconducting  order parameter at site $a$.
We discuss the actual construction of such elements in Sec.
\ref{Implementations}.
Due to this $2 \pi/n$-periodicity each link has $n$ degenerate ground states
characterized by an integer, $u_{ab}=0,\dots ,n-1$ which is defined by
%
%
\begin{equation}
\Delta\phi_{ab}=\phi_a-\phi_b= 2\pi u_{ab}/n
\mbox{.}
\label{Delta_phi_ab}
\end{equation}
%
%
We would like to stress that this type of system exhibits a {\em local}
$\nbZ_n$ symmetry defined by $\phi_a \rightarrow \phi_a + 2\pi m_a/n$
with integers $m_a$.
Since we are dealing with quantum systems, it is also natural to interpret
this symmetry in a basis which diagonalizes the local particle number operators
conjugated to the phase operators. This symmetry then simply means that the
particle number on each lattice site is conserved modulo $n$. An
explicit realization
for  the $n=2$ case has been recently proposed in Refs.
\cite{Doucot2002,Ioffe2002}.

\begin{figure}[ht]
\includegraphics[width=2.0in]{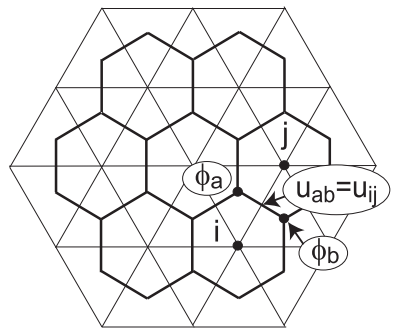}
\caption{Schematics of the Josephson-junction array equivalent to the Abelian
theory. Physical variables (superconducting phases) are defined on the hexagonal
lattice $\Lambda$: $\phi_a$ are phases of the islands while $u_{ab}$ describe
the phase differences. Each bond represents a superconducting element which
is $2\pi/n$ periodic in the phase difference $\phi_a-\phi_b$. The state of
each bond is described by the discrete variable  $u_{ab}$ according to Eq. (\ref{Delta_phi_ab}).
Alternatively, the state of the bond can be described by the variable $u_{ij}$ defined on
the links of the dual (triangular) lattice $\Lambda_d$.
}
\label{Lattice}
\end{figure}

In real systems, charging energies induce quantum fluctuations of the
phase variables.
Tunneling processes between the various degenerate
classical ground states discussed above appear and usually lift completely this
classical degeneracy. Physical properties in the low-energy subspace spanned by
these classical ground states are then well described by the
following Hamiltonian:
%
%
\begin{equation}
H_{sites} = - \sum_{a\in \Lambda} \sum_{m=0}^{n-1}  r(m) \: {\cal
T}_{a}(m)=\sum_a H_a
\mbox{,}
\label{H_Abelian}
\end{equation}
%
%
where $\Lambda$ denotes the physical lattice. In order to give a
concrete illustration, we work on
the geometry displayed in Fig. \ref{Lattice} where $\Lambda$ is the hexagonal
lattice and its dual
$\Lambda_d$ is the triangular one.
Here, the operator ${\cal T}_{a}(m)$ is defined by
%
%
\begin{equation}
{\cal T}_{a}(m) |\phi\rangle_a= |\phi +2\pi m/ n \rangle_a
\mbox{,}
\end{equation}
%
%
and $r(m)$ is a tunneling amplitude which is expected to be maximal for
$m \equiv \pm 1 [ n ]$.
%
%
%
%
In this case, the Hamiltonian $H_a$ is easily diagonalizable since
its eigenstates are
``plane waves"
%
%
\begin{equation}
|l\rangle_a={1 \over \sqrt{n}} \sum_{m=0}^{n-1} \exp({i2 \pi l  m/n})|\phi +
2\pi m/ n \rangle_a
\mbox{,}
\end{equation}
%
%
and its eigenenergies are given by $E(l)=-2r\cos(2\pi l/n)$ for
$l=0,\dots ,n-1$. Here, we have chosen tunneling amplitudes
$r(m)$ equal to $r$ for $m = \pm 1$ and $0$ otherwise.
Thus the ground state $|{\cal G} \rangle =\otimes_a |{\cal G}\rangle_a$ is
given by a configuration such that $l=0$ for all $a$.

Although this model is very easy to solve since it only involves
single-site operators which mutually commute, it is very fruitful to
map it onto a genuine
local gauge theory also based on the $\nbZ_n$ group. Indeed, this
mapping provides a
convenient way to treat physical external perturbations which are
coupled to link variables.
This construction is based on a description of physical states in terms of the
link variables $u_{ab}$. The crucial observation is that these
variables are not
independent as soon as there are loops on the lattice. More
precisely, the sum of
the integers $u_{ab}$'s along any oriented loop is constrained to
vanish modulo $n$.
A natural basis for our system is given by states of the form
$|\Psi \rangle = \otimes_{ab} |u \rangle_{ab}$, where the $u_{ab}$'s
satisfy the constraint.
In the following, we shall refer to the  description  in terms of the
$\phi_a$'s as ``site"
representation, and to the latter (in terms of the $u_{ab}$'s) as the
``link" representation.
This condition on the $u_{ab}$'s can be easily interpreted as the
Gauss law for an Abelian
gauge theory on the dual lattice. To show this, it is convenient to
change the basis via the
Fourier transform
%
%
\begin{equation}
|p \rangle_{ab} = {1\over \sqrt{n}}\sum_{u=0}^{n-1} \exp(i 2\pi p u
/n) |u \rangle_{ab}
\mbox{,}
\label{p_ab}
\end{equation}
%
%
where $p$ is an integer chosen modulo $n$.
In this basis, a state $|u \rangle_{ab}$ is an eigenvector for the translation
operators ${\cal T}^{(p)}_{ab}(p)$ with the eigenvalue $\exp (i 2\pi
p \, u/n )$ which provides a simple implementation of these operators
in terms of the original fields. 
Thereafter, we will use the notation ${\cal T}^{(x)}_{ab}(y)$ for the
translation operator
defined by
%
%
\begin{equation}
{\cal T}^{(x)}_{ab}(y) | x \rangle_{ab} = |x +y \rangle_{ab}
\mbox{,}
\end{equation}
%
%
for $x=u,p$. Note that
$[({\cal T}^{(p)}_{ab} (p) ]^n={\cal T}^{(p)}_{ab} (n
p)={\bf 1}$, so that  the
group composed of the
${\cal T}^{(p)}_{ab}$'s is also $\nbZ_n$. The constraint
%
%
\begin{equation}
\sum_{plaq.} u_{ab} \equiv 0 [n] \Leftrightarrow \prod_{plaq.} \exp
(i 2\pi \, u_{ab}/n)=1
\mbox{,}
\label{constraint1}
\end{equation}
%
%
for any elementary plaquette is equivalent to requiring the
invariance of physical states under
the action of the operator $\prod_{plaq.} {\cal T}^{(p)}_{ab}(1)$
which shifts all the
$p_{ab}$'s around the plaquette by 1.

The last stage of the mapping onto a $\nbZ_n$ gauge theory
consists in expressing everything on the dual lattice. Any link $ab$
on $\Lambda$
separates two plaquettes which correspond to sites $i$ and $j$ of $\Lambda_d$.
The oriented links of $\Lambda$ and $\Lambda_d$ are thus in one to
one correspondence.
It is then possible to transform the $p_{ab}$'s into the
$p_{ij}$'s, and similarly
the ${\cal T}_{ab}$'s into the ${\cal T}_{ij}$'s.  We then recover
the Gauss law written as
%
%
\begin{equation}
   \prod_{j(i)} {\cal T}^{(p)}_{ij}(1) |\Psi \rangle = |\Psi \rangle
\mbox{.}
\label{constraint2}
\end{equation}
%
%
In other words, invariance of physical states under $\prod_{plaq.}
{\cal T}^{(p)}_{ab}(1) $
is equivalent to the requirement of invariance under the transformations
$p_{ij} \rightarrow p_{ij}+l_i-l_j$ for any pairs of neighboring
sites $i$ and $j$ belonging
to $\Lambda_d$, and any integers $l_i$ and $l_j$.
This is exactly the equivalent of condition (\ref{GaugeInv}) for a
gauge theory associated with
the group $\nbZ_n$ on the dual lattice where the gauge ``fields" are
the $p_{ij}$'s.

The simplest Hamiltonian invariant under these transformations reads
%
%
\begin{equation}
H =- \sum_{a} \sum_{u=0}^{n-1} r(u) \prod_{b(a)}  {\cal T}^{(u)}_{ab}(u) +
\sum_{ab} F(\hat{u}_{ab})
\mbox{.}
\label{H_Abelian1}
\end{equation}
%
%
The first term gives quantum dynamics to the links and is equivalent
to $H_{sites}$.  The second term $F(\hat{u}_{ab})$ describes 
perturbations which lift the degeneracy between the classical ground
states. In the physical implementations in Josephson-junction arrays
discussed in more detail below, it is usually due to the unprecise value of
the flux through the elementary plaquette. For instance, in the realization 
\cite{Ioffe2002} of  $\nbZ_2$ theory it is equal to $u_{ab} E_J    \delta \Phi/\Phi_0 $. 
Generally, this term breaks the  invariance  $\phi_a \rightarrow \phi_a + 2\pi m_a/n$
whereas it preserves the local gauge symmetry $p_{ij} \rightarrow
p_{ij}+l_i-l_j$ since it is diagonal in the $|u\rangle$ representation.

The Hamiltonian $H$ can also be written as
%
%
\begin{equation}
H=- \sum_{ijk}  \tilde r(\hat{p}_{ij}+ \hat{p}_{jk} +\hat{p}_{ki}) +
{1 \over n} \sum_{ij}  \sum_{p=0}^{n-1} \tilde F(p) {\cal T}^{(p)}_{ij}(p)
\mbox{,}
\label{H_Abelian2}
\end{equation}
%
%
where $\tilde r$ and $\tilde F$ denote the Fourier transform defined by
%
%
\begin{equation}
\tilde K(p)=\sum_{u=0}^{n-1} \exp ( {-i 2\pi p u /n }) K(u)
\mbox{,}
\end{equation}
%
%
for any function $K$.
As usual, for a lattice gauge theory, the first term involves the
``magnetic" flux
$(\hat{p}_{ij}+ \hat{p}_{jk} +\hat{p}_{ki})$ on the plaquettes $ijk$ of
$\Lambda_d$, whereas the second term is off diagonal in the $| p
\rangle$ representation
and is associated with ``electric" fields.

In the absence of this latter term, the model can be solved and its
ground state, in the
$| u \rangle$ representation, reads
%
%
\begin{eqnarray}
|{\cal G} \rangle
&=& \left[\prod_{plaq.} \delta \left(\sum_{(ab)\in plaq.} u_{ab}
\right) \right]
\otimes_{ab} \sum_{u=0}^{n-1} |u \rangle_{ab}
\label{G_Abelian0}\\
&=&\left[  \prod_{i} \delta \left( \sum_{j(i)} u_{ij} \right) \right]
\otimes_{ij}
\sum_{u=0}^{n-1} |u \rangle_{ij}
\mbox{,}
\label{G_Abelian1}
\end{eqnarray}
%
%
where the first factor ensures the constraint (\ref{constraint1}).
The ground state can also
be simply written in the $| p \rangle$ representation on the dual lattice
%
%
\begin{equation}
|{\cal G} \rangle= \prod_{i} {\cal P}_i
\otimes_{ij} |0 \rangle_{ij}
\mbox{,}
\label{G_Abelian2}
\end{equation}
%
%
where
%
%
\begin{equation}
{\cal P}_i=\sum_{p=0}^{n-1} \prod_{j(i)} {\cal T}_{ij}^{(p)}(p)
\mbox{,}
\end{equation}
%
%
is the projector on the gauge-invariant subspace defined by Eq. (\ref{constraint2}), and
where $|0 \rangle$ stands for the states $|p=0 \rangle$.

\begin{figure}[ht]
\includegraphics[width=2.6in]{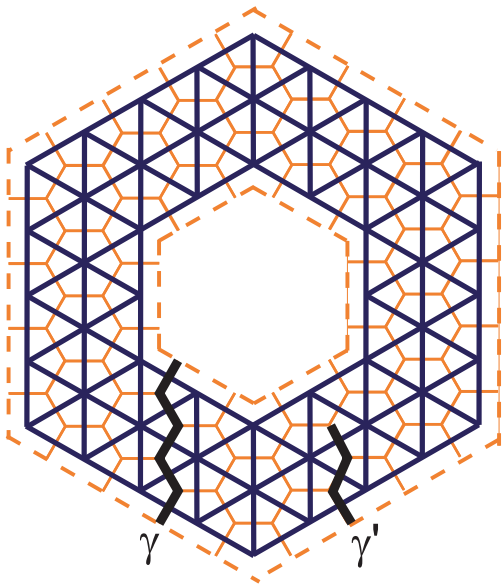}
\caption{
The global topological invariant characterizing the ground state of the system
is defined as a product of operators along contour which joins inner
and outer boundaries
(left bold line). Excitations (fluxons) correspond to the string of
operators that are
illustrated by the right bold line.
}
\label{Lattice2}
\end{figure}

The elementary excitations that we shall call {\em fluxons}
correspond to plaquettes on
$\Lambda_d$ for which the corresponding flux is nonvanishing. Such
an excitation is
produced by an open string of operators
%
%
\begin{equation}
\hat f=\prod_{(ab)\in \gamma} \exp(i2 \pi {\hat u}_{ab} /n)=
\prod_{(ab)\in \gamma} {\cal T}_{ab}^{(p)}(1)
\mbox{,}
\label{fluxon_AB}
\end{equation}
%
%
where the contour $\gamma$ goes over the links of the physical
lattice $\Lambda$ that is, over the centers of plaquettes of $\Lambda_d$ on which the
gauge field $p_{ij}$ is defined (see Fig. \ref{Lattice2}). Assuming $r(m)=r$ if $m=\pm 1$
and $0$ otherwise, these elementary excitations have energy $E=2 r[1-\cos(2\pi/n)]$.  
Physically, in the system of Josephson junctions these excitations  carry
electrical charge $2e$.
We emphasize, however, that in terms of the equivalent gauge theory these
excitations carry no charge,
they describe the fluctuations of the gauge field without  matter.
%
%
%
%

Matter appears in this effective theory if one allows sites of the
dual lattice where  constraint
(\ref{constraint2}) is violated. In terms of the physical
Josephson-junction system, it
means that we  have plaquettes of the physical lattice $\Lambda$
where the sum of the
superconducting phases is  different from $0$, that is vortices in
the superconducting
phase. Notice that due to
$2\pi/n$  periodicity these vortices carry $1/n$ flux quanta. Thus,
it is natural to introduce
sites $i$ where $\sum_{j(i)} u_{ij} \equiv 1  [n] $. In the gauge
field language, it implies
that gauge transformation at these  sites changes the wave function
of the gauge field
according to
%
%
\begin{equation}
\prod_{j(i)} {\cal T}^{(p)}_{ij}(p) | \Psi\rangle= \exp ( {i 2\pi p
/n }) | \Psi\rangle
\mbox{,}
\label{AGInvWithMatter}
\end{equation}
%
%
meaning that, at these sites, there is a pseudocharge $1/n$.
In order to create such an excitation above the ground state, one
needs to apply operators
${\cal T}_{ij}^{(u)}(1)$ along an open string on the dual lattice $\Lambda_d$.
Because of the presence of this string, it is in general necessary to
consider globally
neutral charge configurations. This arises because the product of all
the generators
$\prod_{j(i)} {\cal T}_{ij}^{(p)}$ over all sites $i$ is the identity
operator, or
equivalently because global gauge transformations do not modify the
link variables $p_{ij}$.
When such a pseudocharge moves around a fluxon, it undergoes an
Aharonov-Bohm effect with a
dephasing  $\exp(i 2 \pi /n)$. This is a natural result for a
Josephson-junction array where a fluxon carries a physical charge
$2e$ and where a
pseudocharge  is a superconducting vortex with flux quanta $\Phi_0/n$.
The same result is obtained if we move a fluxon around a fixed
pseudocharge as in the
Aharonov-Casher effect. A more detailed discussion of this phenomenon
is given in the context of non-Abelian gauge theories in Sec. \ref{DNAGT}.

The classification of the excited states as fluxons and pseudocharges
is meaningful
if these charges are rare. This is indeed the case if the energy needed to
create a vortex of the superconducting order parameter (which becomes
pseudocharge in the
gauge theory language) is large compared with the Coulomb energy and also with
the energy scale $r$ of phase fluctuations. Consider qualitatively the
opposite case in which the tunneling amplitude $r$ is large compared to the
potential energy of vortices and to their kinetic energy $t$.
In a physical Josephson-junction array, this can be realized by
constructing the links
of the lattice from rhombi \cite{Ioffe2002} connected in series to an
additional  weak Josephson junction such that its Josephson energy
$\epsilon_J$ and its
charging energy $\epsilon_C$ satisfy
$\epsilon_J < r$ and $\epsilon_J \lesssim \epsilon_C$ \cite{Doucot2003}.
In these conditions all energy scales associated with weak junctions
($\epsilon_C$, $\epsilon_J$, vortex tunneling amplitude $t$) are
small compared to $r$.
Thus, in the leading approximation one can neglect the dynamics of the
weak junctions and their contribution to the potential energy. The
leading process remains the simultaneous flip of three adjacent rhombi
described by the first term of the Hamiltonian (\ref{H_Abelian1}).
The condition $\sum_{j(i)} u_{ij} \equiv 0 [n]$ no longer holds
because weak links favor the
spontaneous excitation of vortices. Instead, it is replaced by the
condition that all
low-energy states have to minimize the kinetic energy part of the Hamiltonian
(\ref{H_Abelian1}), i.e., to satisfy
%
%
\begin{equation}
\prod_{b(a)}{\cal T}_{ab}^{(u)}(1) | \psi\rangle = | \psi\rangle.
\label{Gauss_Ins}
\end{equation}
%
%
This condition can be viewed as a Gauss law imposed on the sites of the
physical lattice under the gauge group generated by ${\cal T}^{(u)}_{ab}$
[cf. Eq. (\ref{constraint2})]. In these conditions, the physical
array is expected
to be in the insulating regime. It is therefore natural that the discrete
version (\ref{Gauss_Ins}) of the QED gauge invariance (\ref{Gauss})
should be restored.
Formally, the states satisfying the gauge-invariance condition
(\ref{Gauss_Ins}) are described by the same theory as the states
satisfying Eq. (\ref{constraint2}) but the physical meaning of all quantities
is reversed. Fluxons correspond now to the fractional vortices in the
superconducting
phase, while pseudocharges become the physical charges
of the insulating array (modulo $n$).
Finally, note that the degenerate states
appearing in arrays with a nontrivial topology are distinguished by
the product of ${\cal T}_{ij}^{(u)}(1)$ operators taken
over paths defined on the dual lattice.

To conclude this section, we emphasize that when the gauge group is Abelian, there
is a duality between the theory defined on the ``real lattice" and the
theory defined on its dual. This correspondence is established via the Fourier
transform which relates the variables $u$ and $p$ defined on the real and on the dual
lattices, respectively. In addition, when both structures are isomorphic, the gauge
theory is self-dual, but it is not the case for an arbitrary system. In this
sense, $\nbZ_{n}$ lattice gauge theories on an hexagonal lattice are not
self-dual. However, the important characteristic required for the physical
operations discussed in our paper is not self-duality but duality which is
always satisfied.

%
%
%

%
%

\section{Discrete non-Abelian gauge theories}
\label{DNAGT}

%
%
%
In this section, we consider the case of a finite non-Abelian gauge
group $G$. As already discussed in Sec. \ref{CGT}, it then becomes necessary to consider
models resulting from the gauge-invariance requirement and the kinematical Gauss law
constraint separately.
%
%
%

\subsection{Gauge-invariant model}
\label{DNAGT1}

%
%
%

The requirement of local gauge invariance simply means that the
wave function $|\Psi\rangle$
satisfies $U_{\{g\}}  |\Psi \rangle=|\Psi \rangle$ for any
configuration of the $g_i$'s,
where  $U_g$ is defined by
%
%
\begin{equation}
U_g=\prod_i \prod_{j(i)} {\cal T}_{ij}(g_i)
\mbox{.}
\end{equation}
%
%
Here, we have adopted multiplicative notation for the group law and
we have introduced the left ``translation" operator ${\cal T}_{ij}
$ defined by
%
%
\begin{equation}
{\cal T}_{ij}(h) |g \rangle_{ij} = |h \, g \rangle_{ij}
\mbox{.}
\end{equation}
%
%
As previously, link variables are oriented in the sense that $|g
\rangle_{ij}=|g^{-1}
\rangle_{ji}$.
The Hamiltonian invariant under those transformations has the form similar to
Eq. (\ref{H_Abelian2}):
%
%
\begin{equation}
H=- \sum_{ijk}  \tilde r(\hat{g}_{ij} \hat{g}_{jk} \hat{g}_{ki}) +
\sum_{ij}  \sum_g \tilde F(g) {\cal T}_{ij}(g)
\mbox{,}
\label{H_NA}
\end{equation}
%
%
where $\tilde{r}$ and $\tilde{F}$ are now required to be invariant under
inner group  transformations $g \rightarrow hgh^{-1}$ for any $h$ in $G$.
As before, $\tilde r$ is associated with the energy cost of a
fluxon-type excitation.
Note that for a continuous group, the $g_{ij}$'s are naturally
represented by matrices and
$\tilde r$ is often chosen as the trace function. By contrast, for a
discrete group, we choose
here the simplest case for which $\tilde r(e)=r$ and $0$ otherwise, a
choice that differs from
previously where we had $\tilde r(p)=2 r \cos(2\pi p/n)$.
As in the Abelian case, this model can be solved in the limit
$\tilde F \rightarrow 0$, where one can construct the ground state explicitly
%
%
\begin{equation}
|{\cal G} \rangle = \prod_{i} {\cal P}_i \otimes_{ij} |e \rangle_{ij}
\mbox{,}
\label{G_NA}
\end{equation}
%
%
where
%
%
\begin{equation}
{\cal P}_i= \sum_{h} \prod_{j(i)} {\cal T}_{ij}(h)
\mbox{,}
\end{equation}
%
%
is the projector on the gauge-invariant subspace. Note that this expression of
the ground state is similar to the one given in Eq. (\ref{G_Abelian2}) for
the Abelian $\nbZ_n$
group.

As in the Abelian case, the excitations are obtained by changing the
value of the $\tilde r$ on
one plaquette. Nevertheless, the non-Abelian nature of $G$ prevents us
to write down a
compact form for the fluxon creation operator $\hat f$ which would generalize
expression (\ref{fluxon_AB}).
Instead, we first consider the state
$\prod_{ (k l) \in \gamma} {\cal T}_{kl} (g) \otimes_{ij} |e \rangle_{ij}$
where the product goes over the links that are intersected by contour $\gamma$
drawn on $\Lambda$ that begins at the boundary of the lattice and
ends at the excited
plaquette. Then, one projects it onto the allowed Hilbert space getting
%
%
\begin{equation}
|f\rangle = \prod_{i} {\cal P}_i \prod_{ (k l) \in \gamma} {\cal T}_{kl} (g)
\otimes_{ij} |e \rangle_{ij}
\mbox{.}
\label{f_NA}
\end{equation}
%
%
Since the projection operator $\prod_{i} {\cal P}_i$ is gauge-invariant, one can
insert any gauge transformation operator between it  and
$\prod_{ (k l) \in \gamma} {\cal T}_{kl} (g)$ without changing $|f\rangle$.
Inserting the global gauge transformation with a group element $h$, we
transform the string of ${\cal T}_{kl} (g)$ operators into the string
of ${\cal T}_{kl} (g')$
with  $g'=h^{-1}g h$. Thus, all group elements that belong to the
same conjugacy class result
in the  same excitation with energy $E= r$.

It is useful to give a more intrinsic description of these fluxons by
considering generalized
magnetic fluxes along closed paths. For such a path $\gamma_c$, the
generalized magnetic flux
is defined as $\Phi({\gamma_c})=\prod_{(ij) \in {\gamma_c}} g_{ij}$.
In the absence of fluxon,
one has $\Phi({\gamma_c})=e$ independently of $\gamma_c$. In the
fluxon state $|f \rangle$,
one has $\Phi({\gamma_c})=e$ only if $\gamma_c$ does not encircle the
plaquette on which the
excitation is localized. If the path encloses this plaquette once,
then $\Phi({\gamma_c})$
belongs to the conjugacy class of $g$, where $g$ is the same group
element as in Eq. (\ref{f_NA})
provided the orientation of $\gamma_c$ coincides with the one of the
string operator used to
build $|f \rangle$. Let us mention three important properties of
these generalized fluxes. For
this purpose, we need to consider closed paths $\gamma_c$ which begin
and end at a fixed
arbitrary site $O$. In this case, we shall say that $O$ is the origin
of $\gamma_c$ or that
$\gamma_c$ is based at site $O$. We have then the following
%
%
\begin{enumerate}

\item If $\gamma_c$ and $\gamma_c'$ are both based at site $O$, and
if one may deform
$\gamma_c$ into $\gamma_c'$ without going through any localized fluxon, then
$\Phi({\gamma_c})=\Phi({\gamma_c'})$.
\label{prop1}

\item If $\gamma_c$ and $\gamma_c'$ involve the same closed loop but
with different origins
$O$ and $O'$, then $\Phi({\gamma_c})$ and $\Phi({\gamma_c'})$ belong
to the same conjugacy class.
\label{prop2}

\item $\Phi({\gamma_c})$ is not gauge-invariant in general, but its
conjugacy class is gauge-invariant.
\label{prop3}

\end {enumerate}
%
%

\begin{figure}[ht]
\includegraphics[width=2.4in]{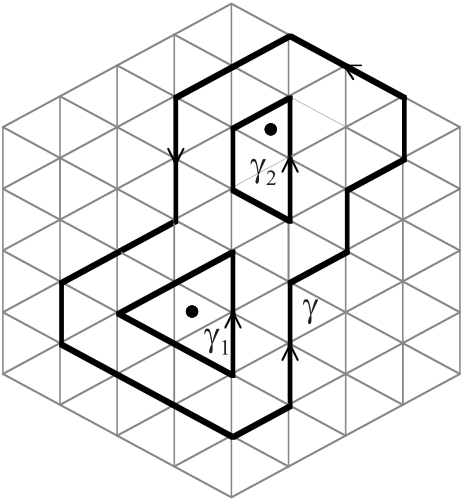}
\caption{
The closed paths $\gamma_1$ and $\gamma_2$ encircle only one fluxon,
whereas $\gamma$ encircles
both. The conjugacy classes of $\Phi({\gamma_1})$, $\Phi({\gamma_2})$, and
$\Phi({\gamma})$ are gauge-invariant and are also invariant under
``smooth" deformation of the
paths. However, $\Phi({\gamma})$ is not conjugated to
$\Phi({\gamma_1}) \Phi({\gamma_2})$, except if all the paths have the
same origin.
}
\label{TwoFluxons}
\end{figure}

For two or more fluxons, the quantum state of the pure gauge theory
is no longer determined by the knowledge of the conjugacy class
associated with each individual
fluxon. This arises because if $g_1$ is conjugated to $g'_1$ and
$g_2$ to $g'_2$, then $g_1
g_2$ is not always conjugated to $g'_1 g'_2$.
Although all these conclusions are valid for an arbitrary group
$G$, we shall explicitly
consider here only the simplest non-Abelian group $D_n$ $(n\geq 3)$
which contains
$n$ rotations $\zeta^k$ ($k = 0,\ldots ,n-1$) and $n$ rotations combined
with reflections $\tau \zeta^k$. The two generators $\zeta$ and $\tau$
satisfy the three relations: $\zeta^{n}=e$, $\tau^{2}=e$, and
$\tau \zeta = \zeta^{-1} \tau$, where $e$ is the unit element of the group.
For simplicity, let us choose the group $D_3$ which contains six elements and
three conjugacy classes:  $\{e\}$, $\{\zeta, \zeta^{2}\}$, and
$\{\tau,\tau \zeta, \tau \zeta^{2}\}$.
If we take $g_1=g'_1=g_2=\tau$ and
$g'_2=\tau \zeta$, then $g_1g_2=e$ and $g'_1g'_2=\zeta$ belong to
different conjugacy
classes.
A complete description of the two-fluxon state requires then
to specify the conjugacy
classes of the fluxes associated with three sets of paths on
$\Lambda_d$: the paths which
encircle each fluxon only once without going around the other, and
the paths which encircle
both fluxons only once (see Fig. \ref{TwoFluxons}).
%
%
%
%

Here, we have used the properties (\ref{prop1}) and (\ref{prop2}) of
general fluxes given above. The non-Abelian nature of $G$ induces
then a notion of
nonlocality when we combine several fluxons. In this sense, these
objects do not behave as
ordinary independent quasiparticles when we consider their global
quantum statistical properties \cite{Propitius1999}.

%
%
\begin{figure}[htb]
\includegraphics[width=2.4in]{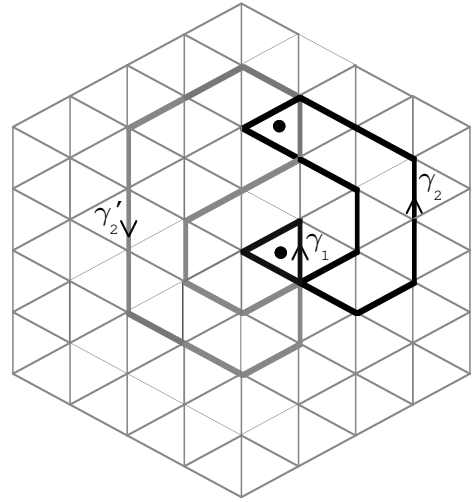}
\caption{
The path $\gamma'_{2}$ is equivalent to
$\gamma_{1}^{-1}\gamma_{2}\gamma_{1}$ which is not in the same
class as $\gamma_{2}$ as this figure illustrates.
Consequently, the fluxes $\Phi(\gamma_2)$ and $\Phi(\gamma'_{2})$
belong to the same conjugacy class, but they are in general different,
in spite of the fact that $\gamma_{2}$ and $\gamma'_{2}$ have the same origin
$O$, that they both wind exactly once counterclockwise around fluxon 2,
and do not wind around fluxon 1.
}
\label{TwoFluxonsA}
\end{figure}
%
%

More precisely, the state of the system is characterized by the
values of flux calculated for a set of closed paths that begin
and end at a given site $O$ of the lattice and avoid the fluxon
positions. The value of flux is the same for all topologically
equivalent paths (with the same origin), i.e., for the paths
that can be smoothly deformed into each other without going over
the fluxon positions.  The set of topologically nonequivalent
paths has a natural group structure (provided by path concatenation),
called the fundamental group of the lattice with $n$ plaquettes removed,
we shall denote it by $\pi_{1}(n)$.
As is well known ~\cite{Mermin1979}, $\pi_{1}(n)$ is generated by the
equivalence classes of $n$ paths, each of which winds exactly once
around only one particular fluxon. But for $n \ge 2$, this group is
non-Abelian as illustrated in Fig. \ref{TwoFluxonsA}. Notice
that specifying a set of winding numbers for a path with respect to each
fluxon does {\em not} determine the total flux along this path which
depends on the order in which the path encircles the fluxons.
Formally,  any configuration of link variables $g_{ij}$ with $n$
fluxons establishes a (non-Abelian) group homomorphism from
$\pi_{1}(n)$ to $G$, which sends the equivalence class of a path $\gamma$
based at $O$ into $\Phi(\gamma)$. Clearly, $\Phi(\gamma_{1}\gamma_{2})=
\Phi(\gamma_{1})\Phi(\gamma_{2})$. In the simpler case of Abelian $G$,
this subtle effect due to the non-Abelian nature of $\pi_{1}(n)$
is washed out and fluxes do depend only on the set of winding numbers
and {\em not} on their actual sequence along a path. Finally, we wish to
characterize {\em gauge-invariant} states with $n$ fluxons. As already
noted, a gauge transformation $g'_{ij}=h_{i}g_{ij}h^{-1}_{j}$ transforms
fluxes $\Phi (\gamma)$ into $\Phi ' (\gamma) = h_{0} \Phi (\gamma) h_{0}^{-1}$.
Thus, equivalence classes of group homomorphisms from $\pi_{1}(n)$ to $G$
are in one to one correspondence with the gauge-invariant states for $n$
fluxons.

Let us illustrate this for $n=2$ and $G=D_{3}$.
If fluxons were described
by group elements, we would expect $6^{2}=36$ states. But this clearly
does not take the gauge-invariance condition into account. If we naively
associate conjugacy classes to fluxons, this yields $3^{2}=9$ states
for two fluxons. It turns out that the correct counting is $11$. We list
below the corresponding equivalence classes of group homomorphisms
from $\pi_{1}(2)$ to $D_{3}$. Since $\pi_{1}(2)$ has two independent generators
$\gamma_{1}$ and $\gamma_{2}$ (see Fig. \ref{TwoFluxonsA}), this
homomorphism is completely determined by the knowledge of
$\Phi(\gamma_{1})$ and $\Phi(\gamma_{2})$.
In these notations, the eleven possible states of two fluxons are:
\begin{equation}
\begin{array}{l}
\Phi_{00}=\{(e,e)\}, \\
\Phi_{01}=\{(e,\zeta),(e,\zeta^{2})\}, \\
\Phi_{10}=\{(\zeta,e),(\zeta^{2},e)\}, \\
\Phi_{02}=\{(e,\tau),(e,\tau\zeta),(e,\tau\zeta^{2})\}, \\
\Phi_{20}=\{(\tau,e),(\tau\zeta,e),(\tau\zeta^{2},e)\}, \\
\Phi_{11}=\{(\zeta,\zeta),(\zeta^{2},\zeta^{2})\}, \\
\Phi_{11}'=\{(\zeta,\zeta^{2}),(\zeta^{2},\zeta)\}, \\
\Phi_{22}=\{(\tau,\tau),(\tau\zeta,\tau\zeta),(\tau\zeta^{2},\tau\zeta ^{2})\}, \\
\Phi_{22}'=\{(\tau,\tau\zeta),(\tau,\tau\zeta^{2}),(\tau\zeta,\tau),
(\tau\zeta,\tau\zeta^{2}),(\tau\zeta^{2},\tau), \\ (\tau\zeta^{2},\tau\zeta)\}, \\
\Phi_{12}=\{(\zeta,\tau),(\zeta,\tau\zeta),(\zeta,\tau\zeta^{2}),
(\zeta^{2},\tau), (\zeta^{2},\tau\zeta), (\zeta^{2},\tau\zeta^{2})\}, \\
\Phi_{21}=\{(\tau,\zeta),(\tau\zeta,\zeta),(\tau\zeta^{2},\zeta),
(\tau,\zeta^{2}),(\tau\zeta,\zeta^{2}),(\tau\zeta^{2},\zeta^{2})\}.
\end{array}
\end{equation}
Note that in this list there are two pairs of \textit{different} two-fluxon
states, $\Phi_{11}$, $\Phi_{11}'$ and $\Phi_{22}$, $\Phi_{22}'$
which are produced by the combination of \textit{nondistinguishable} one-fluxon states, 
namely, states
with fluxes $\Phi_{1}=\{\zeta,\zeta^{2}\}$ and
$\Phi_{2}=\{\tau,\tau\zeta,\tau\zeta^{2}\}$.

As usual, including static local charges in such a model requires to
consider sites which transform differently under the gauge group.
More precisely, these sites can be  classified according to
the representations of the gauge group. For a non-Abelian group,
most irreducible representations are no longer one dimensional.
This leads us to attach a finite-dimensional representation space
spanned by basis vectors $|\alpha \rangle$ to each of these sites.
The corresponding representation of $G$ in this subspace is described
as usual by a set
of matrices $D_{\alpha \beta}(g)$. Now, let us assume that we have
only one such site. The
global wave function may be written as
$|\Psi \rangle = \sum_{\alpha} |\alpha \rangle \otimes|\Psi_{\alpha}
\rangle  $. In this
slightly generalized setting, the  local gauge-invariance
constraint becomes
$|\Psi_\alpha \rangle = \sum_{\beta} D_{\alpha \beta} (g)
U_{g}|\Psi_\beta \rangle$, if the
group element $g$ is attached to a site carrying a nontrivial representation.
For the $D_n$ group that we consider here, the most
interesting cases are obtained from the two-dimensional irreducible
representation of the group.
The wave function of the ground state satisfying this constraint at
one site $o$, occupied by
a pseudocharge in the $D(g)$ representation may be written explicitly as
%
%
\begin{equation}
|{\cal G}\rangle = \left( \prod_{i \neq o} {\cal P}_i \right) \tilde{\cal P}_o
|\alpha \rangle \otimes_{ij} |e \rangle_{ij}
\mbox{,}
\label{C_NA}
\end{equation}
%
%
where $|\alpha \rangle$ is any vector in the charge Hilbert space, and
$\tilde{\cal P}_o=  \sum_{h} D(h) \otimes \prod_{j(o)} {\cal T}_{oj}(h)$.

\begin{figure}[ht]
\includegraphics[width=2.4in]{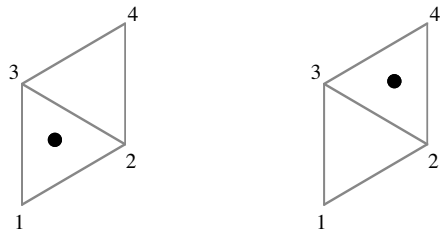}
\caption{
Fluxon moving from plaquette (123) to plaquette (234) involves a
modification of $g_{23}$. In the
initial state (left), $g_{23}^{init.}=g_{24}g_{43}$, and in the final state
(right) $g_{23}^{fin.}=g_{21}g_{13}$.
The operator which transforms the former into the latter state is
${\cal M}_{23}$.
}
\label{onefluxon}
\end{figure}

As for the Abelian case, it is interesting to discuss the situation
where a pseudocharge moves
around a fluxon and conversely. The state combining a pseudocharge
$|\alpha \rangle$ located at site $i$ and the gauge degrees of freedom in
the state $| \Psi \rangle$ will be denoted by
$|\alpha \rangle_i \otimes | \Psi \rangle$. In a trivial background where
$| \Psi \rangle= \otimes_{ij} | e \rangle_{ij}$, it is natural to
assume that the charge
hopping Hamiltonian $H_t$ acts according to
%
%
\begin{equation}
H_t (|\alpha \rangle_i \otimes | \Psi \rangle) = -t \sum_{j(i)}
|\alpha \rangle_j \otimes |
\Psi
\rangle
\mbox{,}
\label{H_t}
\end{equation}
%
%
where $t$ is a positive hopping amplitude. Requiring
invariance of $H_t$ under local
gauge transformations completely determines the action of $H_t$ for
a nontrivial gauge field
background $| \Psi \rangle$:
%
%
\begin{equation}
H_t (|\alpha \rangle_i \otimes | \Psi \rangle) = -t \sum_{j(i)}
[D(g_{ij}^{-1})|\alpha
\rangle]_j
\otimes |
\Psi
\rangle
\mbox{.}
\end{equation}
%
%
Because of the string involved to build the fluxon $|f \rangle$, it
is clear that a
pseudocharge moving around it experiences a generalized
Aharonov-Bohm effect which is, in
general, nondiagonal in the Hilbert space associated with the internal
degrees of freedom of the
charge.
Conversely, we may consider the case of a fluxon associated with
the conjugacy class of $g$ and moving around a
fixed charge. When the fluxon
moves from one plaquette to an adjacent one, we obtain the final
state by modifying only the
field variable $g_{ij}$  on the link $(ij)$ that has been crossed by
the fluxon.
This defines
an operator ${\cal M}_{ij}$ which is equal to ${\cal T}_{ij}(g')$ where $g'$
is a group element belonging to the same conjugacy class as
$g$. Note that the choice of this representative element inside the
class of $g$ depends on
the actual field configuration. Consider, for instance, the example of two
plaquettes shown in Fig. \ref{onefluxon}.
%
%
%
%

Thanks to this construction, ${\cal M}_{ij}$
commutes with all the local gauge transformations operators
$U_{\{h\}}$ acting on the gauge
degrees of freedom. Let us take an initial state
${\cal P} (|\alpha \rangle_i \otimes | \Psi \rangle)$, where
${\cal P}=\left( \prod_{i \neq o} {\cal P}_i \right) \tilde{\cal P}_o$. We move
the fluxon along a closed path $\gamma_c$
which encircles the origin $o$ where the pseudocharge is located.
This leads to the state:

%
%
\begin{eqnarray}
\left( \prod_{(ij) \in\gamma_c} {\cal M}_{ij}\right){\cal P} (|\alpha
\rangle_o \otimes |
\Psi \rangle) &=& \\
{\cal P} (|\alpha \rangle_o &\otimes& \left(\prod_{(ij) \in\gamma_c}
{\cal M}_{ij}\right)|
\Psi \rangle)
\mbox{.}
\nonumber
\end{eqnarray}
%
%
Since $| \Psi \rangle$ and ${\cal M}_{ij}| \Psi \rangle$ both
describe the same fluxon, there is
a gauge transformation $U_{\{h\}}$ connecting these two states. We
choose ${\{h\}}$ such that
$h_i=e$ for $i$ outside the area enclosed by the fluxon trajectory.
In this case, $h_o$ is
conjugated to the flux $\Phi=g$ associated with the fluxon. Our final
state is then
${\cal P} [D(h_{o}^{-1})|\alpha \rangle_o \otimes |\Psi \rangle]$. We
thus recover the same
non-Abelian transformation after a closed cycle as for the moving
charge situation.

\begin{figure}[htb]
\includegraphics[width=3.0in]{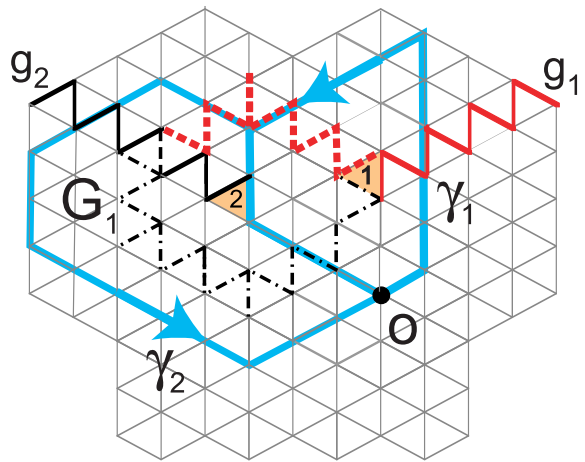}
\caption{
The effect of motion of one fluxon around another.
Two fluxons, $1$ and $2$ carry initial fluxes $\Phi_1=g_1$ and $\Phi_2=g_2$
that are calculated as a product of group elements along contours $\gamma_1$
and $\gamma_2$ beginning and ending at the same point $O$ of the lattice.
These fluxons can be represented as strings of $g_1$ and $g_2$
(full thick bonds) against the background of unit group elements.
When fluxon 1 moves around fluxon 2, it leaves behind a string of group elements
(dashed and dotted bonds): $g_{1}$ initially, until it crosses the $g_{2}$ string,
and afterwards $G_1=g_{2}^{-1}g_{1}g_{2}$.
After this circular motion the fluxes become $\Phi_1=g_{2}^{-1}g_{1}g_{2}$
and $\Phi_2=g_{2}^{-1}g_{1}^{-1}g_{2}g_{1}g_{2}$ respectively.
}
\label{TwoFluxonsB}
\end{figure}

Furthermore, non-Abelian Aharonov-Bohm effects also
appear when a fluxon goes around another fluxon \cite{Bais1980}.
This situation is shown in Fig. \ref{TwoFluxonsB}, where fluxon 1
carrying initially the flux $\Phi_1=g_{1}$ goes around fluxon 2
(with initial flux $\Phi_2=g_{2}$) once counterclockwise. These fluxes
are defined as a product of group variable around contours $\gamma_1$
and $\gamma_2$ starting and ending at point $O$ or any other contours
topologically equivalent to them. The state with these fluxes can be
represented by the configuration of  $g_{ij}$ that contains two strings
against the ``flat", $g_{ij}=e$, background.
As the first fluxon moves around the second one, its string eventually
crosses the string of the static fluxon. After the crossing the string
of the first fluxon changes into the string of $G_1=g_{2}^{-1}g_{1}g_{2}$
elements. The new fluxes $\Phi_{1}$ and $\Phi_{2}$ become
$\Phi_{1}=g_{2}^{-1}g_{1}g_{2}$ and
$\Phi_{2}=g_{2}^{-1}g_{1}^{-1}g_{2}g_{1}g_{2}$. Since
$g_{2}^{-1}g_{1}g_{2}=g_{2}^{-1}g_{1}^{-1}g_{1}g_{1}g_{2}$, this
state can be reduced to the initial state by a \textit{global} gauge transformation
with the group element $g=g_1 g_2$,
\textit{if} there are only two fluxons in the system. But if some other fluxons
are present somewhere else in the system, besides the two already discussed,
they might transform nontrivially under the gauge transformation
with $g$: $\Phi \ne g^{-1} \Phi g $ and in a general case the final state
cannot be reduced to the initial state by the gauge transformation.
This provides another example of nondiagonal Aharonov-Bohm effect, even in
the absence of external charges. These considerations have been generalized
to composite objects of fluxons and
pseudocharges usually called dyons \cite{Propitius1999}. The appropriate
mathematical framework has been developped in terms of representation
theory of the so-called quantum double algebra associated with the group $G$
\cite{Dijkgraaf1990,Bais1992}. We shall not elaborate further on these
interesting topics here.
\bigskip
\bigskip

%
%
%
%

As for Abelian models, the systems with one or several holes
have a degenerate ground state. In the limit of zero off-diagonal
term in Hamiltonian (\ref{H_NA}) the fluxons are trapped forever
in the holes and have zero energy. The degeneracy of the ground
state with one hole is thus equal to the number of different fluxons, i.e., to
the number of conjugacy classes. The operator that distinguishes
these ground states is $\prod_C g$ where contour $C$ encirles the
hole. A small but nonzero off-diagonal term in the Hamiltonian leads
to an exponentially small amplitude for the fluxon escape from the hole.
The operator leading to this escape process is the product of
${\cal T}_{ab}(g)$ along the contour $\gamma$ on lattice $\Lambda$
that connects inner and outer boundaries as shown in Fig.
\ref{Lattice2}. This is the same
operator that moves fluxons around the system, as discussed above. The gap
for the fluxons implies that this process is only virtual and its
amplitude is exponentially small. Therefore, the resulting degeneracy
splitting is also exponentially small. Note that (as discussed above)
the choice of the
group element $g$ in fluxon motion  operator ${\cal T}_{ab}(g)$
depends on the link
variables. The contour
$\gamma$ crosses \textit{once} any contour $C$ and thus changes the
flux trapped inside the hole.
Since these fluxons undergo non-Abelian Aharonov-Bohm effects, they may be used to construct
a complete set of elementary gates for quantum computing, provided the group $G$ is
``sufficiently" non-Abelian. This has been recently analyzed in great detail by Mochon,
first restricting the discussion to states involving only fluxons \cite{Mochon1},  and then
by extending it to the case where pseudocharges are also involved \cite{Mochon2}. This
latter construction is specially interesting for our purpose, since it shows that the
small $D_3$ group (whose practical implementations are discussed in Sec.
\ref{Implementations}) is already capable of universal quantum computation.

%
%
%

\subsection{Kinematical Gauss law model}
\label{DNAGT2}

%
%
%
As explained at the end of Sec. \ref{CGT}, a different model may be
constructed by
assuming a modified Gauss law in which the constraint is diagonal
with respect to the
variables $g_{ij}$. It reads
%
%
\begin{equation}
\prod_{j(i)} g_{ij} = e
\mbox{,}
\label{Gauss_NA}
\end{equation}
%
%
and can be directly compared to Eq. (\ref{ModifiedGauss}). This
constraint allows for a simple interpretation which is motivated by the physical systems discussed in
Secs. \ref{DAGT} and \ref{Implementations}. First, we notice that a constraint
associated with the sites of $\Lambda_d$ is also naturally attached to
the plaquettes on
$\Lambda$. We may therefore also write it as $\prod_{plaq.} g_{ab} = e$.
Second, the constraint that the product of group elements around the
plaquette is equal to $e$
implies that one can define group elements on the sites such that
$g_{ab}=g_a g_b^{-1}$.
Further, these group elements are defined up to a global
transformation $g_a\rightarrow g_a h$.
For the Abelian systems discussed in Sec. \ref{DAGT} the group
elements on the sites of
$\Lambda$ had a transparent physical meaning: they corresponded to
the absolute values of
the phase of the superconducting islands.

The simplest Hamiltonian  preserving the constraint (\ref{Gauss_NA}) is
%
%
\begin{equation}
H =- \sum_{a} \sum_g r(g) \prod_{b(a)}  {\cal T}_{ab}(g) +
\sum_{ab} F(\hat{g}_{ab})
\mbox{.}
\label{H_Gauss_NA}
\end{equation}
%
%
Note the close similarity with the Abelian model defined in (\ref{H_Abelian1}).
However, for a general non-Abelian group, we do not know how to reexpress the
Hamiltonian (\ref{H_Gauss_NA}) in terms of a local gauge theory on
$\Lambda_d$. Indeed,
the key tool in this mapping for the $\nbZ_n$ group was the existence
of the Fourier
transform for functions defined on $\nbZ_n$. Here, we do not see any
obvious way to
generalize this trick to functions defined on a non-Abelian group.
Nevertheless, if $F=0$, the
Hamiltonian (\ref{H_Gauss_NA}) does exhibit a local symmetry based on
the sites of
$\Lambda$: $g_{ab}\rightarrow h_a g_{ab} h_b^{-1}$ provided $r(g)$
depends only on the
conjugacy class of $g$. In this case, the model can be solved for
$F\rightarrow 0$
because, in  this limit, we may interpret the Hamiltonian as a sum of
mutually commuting
terms attached to the sites of $\Lambda$
%
%
\begin{equation}
H_{sites}=- \sum_a \sum_g r(g) {\cal T}_{a}(g) = \sum_a H_a.
\label{H_i}
\end{equation}
%
%
Solution for each site depends somewhat on the structure of the
coefficients $r(g)$. For
the simplest case of positive $r(g)$, the ground state is simply
$|{\cal G}\rangle_a = \sum_g | g \rangle_a$. On the whole lattice,
the ground state is thus
given by: $|{\cal G}\rangle=\otimes_a |{\cal G}\rangle_a$.
This ground state can be also represented in terms of the original
link variables:
%
%
\begin{eqnarray}
|{\cal G} \rangle
&=& \left[\prod_{plaq.} \delta \left(\prod_{(ab) \in plaq.} g_{ab} -
e \right) \right]
\otimes_{ab} \sum_g |g\rangle_{ab}\label{G_Gauss0}\\
&=&\left[\prod_{i} \delta \left(\prod_{j(i)} g_{ij}-e \right) \right]
\otimes_{ij} \sum_g |g\rangle_{ij}
\mbox{.}
\label{G_Gauss}
\end{eqnarray}
%
%
We underline that these expressions are the generalizations of
(\ref{G_Abelian0}) and
(\ref{G_Abelian1}) for an arbitrary non-Abelian group.

This model exhibits local excitations attached to the sites of
$\Lambda$ or equivalently to
the plaquettes of $\Lambda_d$. They are analogous to the Abelian
fluxons constructed in
Eq. (\ref{fluxon_AB}) where the site Hamiltonian $H_a$ in a system of
Josephson junctions
involves charging energies. The Hilbert space attached to the site
$a$ forms the so-called regular representation of $G$ where the group element $h$ is
represented by the left multiplication
operator ${\cal T}_{a}(h)$. As already pointed out, these operators
commute with $H_a$ so
that the eigenstates of $H_a$ belong to stable subspaces for the
regular representation of
$G$. For example, if we choose $G=D_3$, its regular representation is
six dimensional and decomposes into two different one-dimensional
representations and two
isomorphic copies of the two-dimensional representation of $D_3$.
At this stage, we should notice that $H_a$ also commutes with the
representation of $G$ by
right multiplications (the element $h$ being represented by ${\cal
R}(h)$  so that
${\cal R}(h)|g \rangle=|g \,h^{-1}\rangle)$.
Because of this extra symmetry, the spectrum of $H_a$ is composed of two
singlets and one quadruplet.

In fact, it is possible to generate the whole set of eigenstates of
$H$ when $F=0$. Let us choose
a reference state $|\Omega \rangle$ in which the constraint
$\prod_{plaq.} g_{ab}=e$ is satisfied
for all plaquettes. Now, for each site, we consider an eigenfunction
$\phi_a(g)$ of $H_a$ with
energy
$E_a$. More precisely, we require
%
%
\begin{equation}
E_a \phi_a(g)=- \sum_h r(h)  \phi_a(h^{-1}g)
\mbox{.}
\end{equation}
%
%
We can then construct an eigenstate  $|\Psi \rangle$ of $H$ with the
total energy $E=\sum_a E_a$ as follows:
%
%
\begin{equation}
|\Psi \rangle=\prod_a \sum_g \phi_a(g) \prod_{b(a)} {\cal T}_{ab}(g)
|\Omega \rangle
\mbox{.}
\end{equation}
%
%
In the case where $\phi_a(g)$ is independent of $g$, we recover the
ground state
$|{\cal G}\rangle$ given by Eq. (\ref{G_Gauss0}). Introducing the
gauge transformation operators
$U_{\{g\}}$ as before (where the $g$'s now live on the sites of
$\Lambda$), we may also write
%
%
\begin{equation}
|\Psi \rangle=\prod_{a} \sum_{\{g\}}  \phi_a(g_a) U_{\{g\}} |\Omega\rangle
\mbox{.}
\end{equation}
%
%
This shows that using a gauge transformed reference state $U_{\{h\}}
|\Omega\rangle$ instead of
$|\Omega\rangle$ is equivalent to replace the site wave functions
$\phi_a(g)$ by
$\phi_a(g h_a^{-1})$. As already discussed, acting on the $\phi_a$'s
by these right
multiplications produces eigenstates with the same energy. Only sites
$a$ where $\phi_a$ belongs to
a multidimensional representation of $G$ generates new states when a
gauge transformation with
$h_a\neq e$ acts on $|\Omega\rangle$.

This language is specially useful to describe a different type of
excitation, analogous to the
pseudocharge in the gauge-invariant model. They are produced in a
physical system in which the
constraint is modified on a given plaquette so that $\prod_{plaq.}
g_{ab}= u$ on this plaquette.
Note that as it stands, this constraint is not preserved by the
action of the Hamiltonian
(\ref{H_Gauss_NA}), but the conjugacy class of $\prod_{plaq.}g_{ab}$
is preserved for any
plaquette. This modified constraint may be incorporated in the
general construction presented above
provided we change the reference state. For instance, we may consider
a string $\gamma$ on
$\Lambda_d$ which runs from the center of the excited plaquette to
the boundary of the system. Our
previous state $|\Omega\rangle$ is then replaced by
$\prod_{(ab) \in \gamma} {\cal T}_{ab} (u)|\Omega\rangle$. Note that
in the context of
physical implementations, as  already mentioned in Sec. \ref{DAGT},
it is more appropriate
to call these excitations vortices.

As in the gauge-invariant model, it is interesting to study the
appearance of nondiagonal adiabatic
transport of a degenerate fluxon excitation around a pseudocharge.
As before, let us describe the
situation according to both viewpoints where either one excitation is
fixed or the other. For a
mobile pseudocharge, we have to transform adiabatically the
reference state. Starting from
%
%
\begin{equation}
|\Psi \rangle_{in}=\prod_{a} \sum_{\{g\}}  \phi_a(g_a) U_{\{g\}}
\left(\prod_{(ab) \in \gamma} {\cal T}_{ab} (u)|\Omega\rangle \right)
\mbox{,}
\end{equation}
%
%
we obtain after one closed path $\beta$ of the pseudocharge around the fluxon:
%
%
\begin{equation}
|\Psi \rangle_{fin}= \prod_{a}\sum_{\{g\}}  \phi_a(g_a) U_{\{g\}}
\left(\prod_{(ab) \in \gamma+\beta} {\cal T}_{ab}
(u)|\Omega\rangle \right)
\mbox{.}
\end{equation}
%
%
The projection operator $\prod_{a}\sum_{\{g\}}   \phi_a(g_a) U_{\{g\}}$
ensures that the resulting state has exactly the same energy as the
initial one, so that the motion of pseudocharge does not excite fluxons.
To make a closer analogy with the gauge-invariant model of Sec.
\ref{DNAGT1} we note that this wave function can be also written as
%
%
\begin{equation}
|\Psi \rangle_{fin}=
\left(\prod_{(ab) \in \beta} {\cal M}_{ab} \right) |\Psi \rangle_{in}
\mbox{,}
\end{equation}
%
%
where operators ${\cal M}_{ab}$'s that move the pseudocharge are defined
on the links of $\Lambda$ exactly as the ${\cal M}_{ij}$'s were defined on
the links of $\Lambda_d$ in Sec. \ref{DNAGT1}, these operators have the
same effect as ${\cal T}_{ab}(u)$ on the flat ($|\Omega\rangle$) state and
they commute with $\prod_{b(a)}{\cal T}_{ab}(g)$ for any site $a$ and any
group element $g$.
Using the same reasonings as for the gauge-invariant model, we see that
$\prod_{(ab) \in \beta} {\cal M}_{ab}$ acts in the same way as a
gauge transformation $U_{\{h\}}$ on the state $|\Psi \rangle_{in}$.
Since $h$ has to be nontrivial (more precisely conjugated to
$u$) at the site where the fluxon is located,
$|\Psi \rangle_{fin}$ is
degenerate with $|\Psi \rangle_{in}$, but these
two states may be linearly independent if the fluxon is created in a
multidimensional representation of $G$.

Alternatively, consider a motion of the fluxon which is due to
a small hopping term, $\hat t(g_{ab})$ [arising, e.g. from $F$ in Eq.
(\ref{H_Gauss_NA})].
Let us denote by $\phi_f(g)$, the single-site wave function associated
with the fluxon. In the absence of a pseudocharge, the fluxon state at
site $a$ corresponds to
%
%
\begin{equation}
|\Psi \rangle_{a}=\sum_{\{g\}}  \phi_f(g_a) U_{\{g\}} |\Omega\rangle
\mbox{.}
\end{equation}
%
%
Acting on this state with the operator $\hat t(g_{ab})$ produces the
state $|\Psi \rangle_{b}$ provided $\phi_f(g_a) \hat t(g_a g_b^{-1})=\phi_f(g_b)$.
\cite{foot1}.
In the presence of a pseudocharge,
$|\Omega\rangle$ is replaced by $\prod_{(ab) \in \gamma} {\cal T}_{ab}
(u)|\Omega\rangle$. If the link
$(ab)$ is on the string $\gamma$, one has:
%
%
\begin{equation}
\hat t(g_{ab}) |\Psi \rangle_{a}=\sum_{\{g\}} \phi_f(g_b u_{ab}^{-1}
) U_{\{g\}}
\left(\prod_{(ab) \in \gamma} {\cal T}_{ab} (u)|\Omega\rangle \right)
\mbox{.}
\end{equation}
%
%
So, as the fluxon hops from $a$ to $b$, its internal
wave function is modified by a right
multiplication  associated with an element conjugated to $u$ and there
is a complete agreement between both descriptions.

Note the close similarity between these results and the ones
for the gauge theory model, provided we exchange the roles of
pseudocharges and fluxons.
For instance, a pseudocharge in the kinematical Gauss
law model experiences a nondiagonal Aharonov-Bohm effect when another
pseudocharge is moved around it, even in the absence of fluxons.
Notice that although the pseudocharges
in the Gauge theory are usually associated with independent
degrees of freedom, this distinction (with fluxons in kinematical
model) is mostly formal. It completely disappears in the
alternative (albeit less popular) gauge theory formalism in which
one keeps only the dynamical variables $g_{ij}$ on the
links and creates pseudocharges by replacing the projector
$\tilde{\cal P}_o$
in Eq. (\ref{C_NA}) by $\sum_{h}\chi(h)\prod_{j(o)}{\cal T}_{oj}(h)$
where $\chi(h)$ is the character of the representation of $G$
associated with the pseudocharge at charge $o$. This approach
emphasizes the analogy between the two situations and is more
appropriate to describe the implementations discussed in the next section.

In the Abelian case, the system with nontrivial topology, i.e., with
holes, acquires a
degenerate ground state, since a fluxon located inside the holes has
a vanishing energy. Such a
fluxon is created by the operator $\prod_{(ab) \in \gamma} {\cal
T}^{(p)}_{ab}(1)$ where the contour $\gamma$
in $\Lambda$  connects the inner and the outer boundaries. This operator
is also equal to
$\prod_{(ab) \in \gamma} \exp(i 2 \pi \hat u_{ab}/n)$ and is conserved by the
Hamiltonian (\ref{H_Abelian1}).
Its eigenstates correspond to linear combinations of various
degenerate ground states with fluxons
present or absent in the holes. This applies directly to the
non-Abelian model of the present
section.  The degenerate ground states are therefore identified by the value of
$\prod_{(ab) \in \gamma} g_{ab}$.  In the site representation, it is just the
group element $g_{in} g_{out}^{-1}$ where $g_{in}$, $g_{out}$
represent the states
of the inner and outer boundaries respectively.
These arguments hold provided that each boundary is described by a single
group element $g$ i.e., that all sites belonging to it are connected by
nonfluctuating bonds. In the Abelian case discussed in Sec. \ref{DAGT} the
boundary (denoted by the dashed line in Fig. \ref{Lattice2}) is just
a superconducting
wire.

The same arguments as the ones developed in the Abelian case
\cite{Ioffe2002} show
that only nonlocal operators in $g_{ab}$ result in the decoherence of
the topologically
degenerate states. This shows that these degeneracies are protected
from decoherence
provided that no physical operator couples to the state $g_a$ of an
individual site.

%
%
%

\section{Implementations}
\label{Implementations}

%
%
%
As explained in Sec. \ref{DAGT}  the array that implements directly (in a
natural basis)
the gauge-invariance constraint for the Abelian $\nbZ_n$ group is
physically an insulator while
the  superconducting array becomes equivalent to the same gauge
theory only after a dual
transformation. Alternatively, the superconducting array can be
thought of as implementing
directly the modified Gauss law discussed in Sec IV B. In the
non-Abelian case these two
theories are no more dual to each other, so we discuss these cases separately.

The common element of all arrays discussed here is the
Josephson-junction circuit that
is characterized by the phase difference $\Delta \phi$ which can take
(in a classical limit)
$n \geq 2$ discrete values $\Delta \phi = 2\pi m/n$; quantum processes
leading to transitions between these states. The actual construction
of this element for
$n=3$ (that we shall call masu
\cite{foot2} below) is shown in Fig. \ref{Masu}. It is built from $n$
more basic elements consisting of two Josephson junctions connected
in series. The Josephson
energy of this element (as a function of the phase difference $\delta
\phi$ between the ends)
contains all higher harmonics: $V_0(\delta \phi)=-2 E_J |\cos\delta
\phi/2|$. Now, connecting
these basic  elements in parallel as shown in Fig. \ref{Masu} and
applying the magnetic field
that shifts  the relative phases of neighboring circuits by $2\pi/n$
gives the desired $2\pi/n$
periodic  potential. The nonzero charging energy of the junctions
leads to transitions
between these classical states whose amplitudes are calculated in Appendix A.

\subsection{Gauge-invariance implementation}
\label{subA}

\begin{figure}[ht]
\includegraphics[width=3.2in]{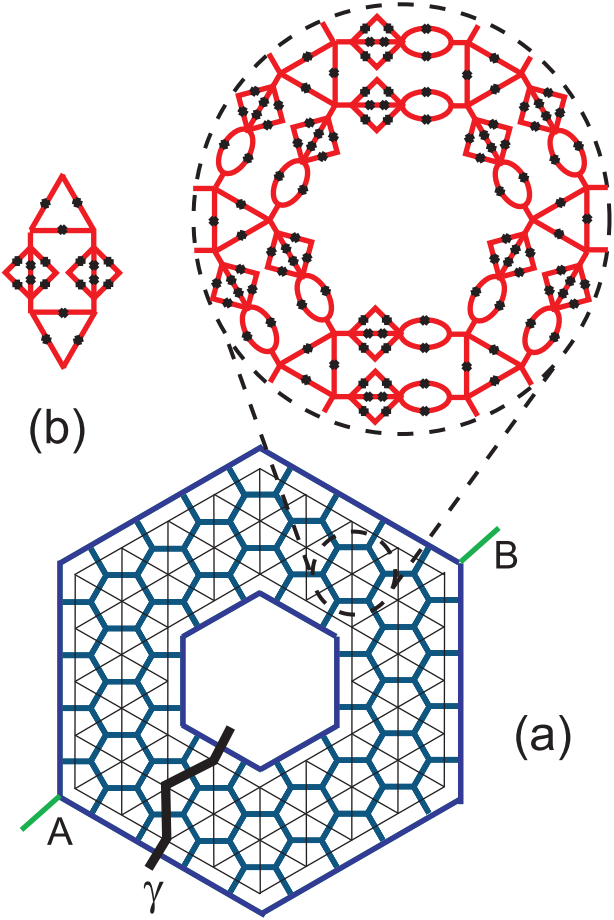}
\caption{
(a) The insulating array implementing non-Abelian insulator. Each
bond of the hexagonal array is built from two circuits, each consisting
of one
masu (rhombus with diagonal) and one weak link (shown as a
Josephson-junction loop with
two junctions). The sites of the array are formed by triangles
frustrated by a
magnetic flux $\Phi_0/2$. 
(b) A simpler construction of \textit{one} bond which
states are in one to one correspondence with the elements of the
permutation group.
A and B denote the leads that can be used to measure the array
properties in the
experiments similar to those discussed in Ref. \cite{Doucot2003}.
}
\label{NonAbelianInsulatorLattice}
\end{figure}

Here we limit ourselves to the gauge theory with the simplest
non-Abelian group $D_3$ which is
also the permutation group of three elements $S_3$. The array
implementing this model is shown
in Fig.
\ref{NonAbelianInsulatorLattice}. We begin by explaining the
qualitative reasoning that leads
to this complicated  construction. In the $S_3$ gauge theory the
state of each link has to be
described by the  permutation of three elements. To implement this in
a Josephson-junction
array, we  construct each bond from two elements (masus) described
by discrete phases
differences
$\Delta \phi_\alpha = 2\pi n /3$ ($\alpha=1,2$). Such a bond can be viewed as a
discrete transformation from the state of one site characterized by
two discrete
$2\pi n /3$ phases to the state of another site; it is described by one of the
nine elements of the $\nbZ_3\times \nbZ_3$ group. In order to
eliminate three extra states
and to get the permutation group, we need to introduce a constraint
requiring that two phases of
each site be different. This is achieved if each site of the lattice
contains an element that
excludes the configurations with equal phases. The simplest of such an
element is a triangle
made of three Josephson junctions frustrated by a magnetic flux
$\Phi_0/2$. The ground
state of this triangle corresponds to the phases $\phi_\alpha=2\pi n_\alpha /3$
with different $n_\alpha$. Thus, the state of two masus attached on
both sides to the
frustrated triangles [as shown in Fig.
\ref{NonAbelianInsulatorLattice}(b)] is in a one to one
correspondence with the elements of $S_3$ group.

%
%
The next step in the construction of the array is to ensure that the
low-energy states satisfy the gauge-invariance constraint.
This is true if the
Hamiltonian is dominated by a large ``kinetic" term which exchanges
two discrete phases on each
triangle because such a process changes the states of all three bonds
connected to the triangle,
further, it is equivalent to the action of a transposition on the
states of these three bonds
\begin{equation}
H_T=-r_0 \sum_{a} \sum_{k=0}^{2} \prod_{b(a)}{\cal T}_{ab}(\tau \zeta^k)
\mbox{,}
\label{H_T}
\end{equation}
where $r_0$ is the tunneling amplitude of this process:
\begin{equation}
r_0 \approx E_J^{3/4} E_C^{1/4} \exp\left(- S_0 \sqrt{\frac{E_J}{E_C}}\right)
\mbox{,}
\label{r_0}
\end{equation}
where $S_0=6.10$ if junctions in triangles have the same Josephson
and charging energies as the ones in masus (see Appendix B). Note
that this numerical value is sensitive to the number of nearest
neighbors of the effective lattice and is minimized for the 
hexagonal lattice. As will be clear from the following, in order
to be described by the gauge theory,  $r_0$ should be sufficiently 
large which is a case easier to achieve for the hexagonal lattice. 

The wave functions minimizing the kinetic energy (\ref{H_T}) are the
ones that satisfy the gauge-invariance constraint
\begin{equation}
\prod_{b(a)}{\cal T}_{ab}(\tau \zeta^k) | \Psi \rangle = | \Psi \rangle
\mbox{,}
\end{equation}
since local transpositions generate the whole $S_3$ gauge group. The
array is faithfully
described by the  gauge theory if $r_0$ sets the largest energy scale
in this problem. This is
{\em not} satisfied  if the bonds of the array are just the ones shown in Fig.
\ref{NonAbelianInsulatorLattice}(b),  because in this case the product
of permutations around the
elementary hexagon is unity and it  is the energy that is needed to
violate this constraint
that sets the largest energy scale in  the array.

In order to decrease the energy scale associated with the
$\prod_{hex.} g_{ab} \neq 1$
configurations,
we need to introduce weak links (shown as loops containing two
Josephson junctions in
Fig. \ref{NonAbelianInsulatorLattice}) characterized by small
Josephson and charging energies
$\epsilon_J \ll \epsilon_C \ll E_C \lesssim E_J$. In order to avoid a
mixture of Josephson
junctions with different parameters $(E_J, E_C)$ in the same array,
we assume that these weak
links are constructed from the two usual Josephson junctions with a
large Josephson energy
$\tilde{E}_J \gtrsim E_J$ and correspondingly small charging energy
$\tilde{E}_C \lesssim E_C$ connected in parallel to form a loop
frustrated by the magnetic
flux. The effective Josephson energy of such a loop is
$\epsilon_J = \tilde{E}_J( \Phi / \Phi_0 -1/2 ) $ and
its charging energy
is $\epsilon_C=\tilde{E}_C/2$. The effect of these small energies on the
tunneling process that corresponds to transposition is weak and can be ignored.
The presence of weak links decreases the energy gap between the six
states of the bond
associated with the permutation group and the remaining three states that
were eliminated
from the low-energy spectrum.
However, as we show below, we can choose the energies of
the weak links, $\epsilon_J$, $\epsilon_C$, so that this gap remains large, but
the amplitude of the tunneling process, $r_0$, remains larger than other
scales in the low-energy theory so that the leading term in the
effective Hamiltonian is
indeed given by Eq. (\ref{H_T}).

The Hilbert space satisfying this constraint is still huge: in the
absence of other
contributions to the Hamiltonian (generated by weak links) all states
satisfying the
gauge-invariance constraint are degenerate. This degeneracy is lifted
when these
subdominant terms are taken into account. Weak links give two types of terms:
they lead to the dynamics of the discrete variables describing the state of
each link and to diagonal potential-energy terms. Let us start with
the latter.  The main
contribution comes in the second order in $\epsilon_J/\epsilon_C$, it
is given by
%
%
\begin{equation}
U(\{u_i\})=- \frac{\epsilon_J^{2}}{2\epsilon_C} \cos\left(
\frac{2\pi}{3} \sum u_i \right)
\mbox{,}
\end{equation}
%
%
where $u_i$ are discrete variables describing the phase differences
on elementary
masus and along the sides of the triangles. This energy is minimized if this sum
is $0$ mo\-dulo $3$, i.e., if the state of the links is described by
the permutation
group ``acting" on the state of triangles. We assume that the scale of
this interaction
is relatively large and comparable to $r_0$, i.e., that 
$ \frac{\epsilon_J^2}{2 \epsilon_C}
\gtrsim r_0$. This interaction does not lift the degeneracy of the
Hilbert space of the
discrete  variables that are described by elements of the permutation
group. This degeneracy is
only lifted by weaker processes which involve longer loops of weak
links around each
hexagon. Such a loop contains six weak links, it results in the
contribution to the potential
energy similar to the first term in Eq. (\ref{H_NA}):
%
%
\begin{equation}
V = \sum_{abcdef} {\cal V}(g_{ab} g_{bc} g_{cd} g_{de} g_{ef} g_{fa})
\mbox{.}
\label{V_NAI}
\end{equation}
%
%
To calculate the value of the interaction ${\cal V}(g)$, we note that
physically
this energy is due to the energy of the Cooper pair that starts from
one corner of
the triangle and moves around the hexagon across the weak links.
Furthermore, if $g$ is
a transposition, the sum of the discrete phases along contours
beginning at two corners
of the triangle is equal to $\pm 2\pi/3$ and one such a sum is $0$,
while if $g$ is a
cyclic permutation, \textit{all} these sums are equal to $\pm
2\pi/3$, so ${\cal V}(e)=0$,
${\cal V}(\tau \zeta^k) = 2 v_0 /3 $, ${\cal V}(\zeta^k) = v_0$,
where coefficient $v_0$ is the strength of the effective Josephson
contact formed by
two parallel loops of six weak junctions with a phase difference
$2\pi/3$ across.
To calculate the value of $v_0$ we note that two parallel loops are
equivalent to one
loop with effective Josephson and charging energies
$\epsilon_J^{eff}=2 \epsilon_J$,
$\epsilon_C^{eff}=\epsilon_C/2$ and use the result of the Appendix C for the
energy of one loop
%
%
\begin{equation}
v_0 = \frac{1}{20} \left( \frac{3 \epsilon_J}{\epsilon_C} \right)^{5}
\epsilon_J.
\label{v_0}
\end{equation}
%
%
We now turn to the dynamical processes which involve phase slips of weak links.
These processes change the state of each link individually, similar
to the second
term in Eq. (\ref{H_NA}). The potential energy involved in such a process
is dominated
by the energy of each masu while the charging energy is due to the
weak junction (because its charging energy is smaller, it dominates the dynamic
term in the action). Thus, the amplitude of such a process is
\begin{equation}
\tilde{F}(g) \approx E_J^{3/4} \epsilon_C^{1/4} \exp(-\tilde{S}), \;\;
\tilde{S} = 2.44 \sqrt{\frac{E_J}{\epsilon_C}}
\mbox{.}
\end{equation}
For small $\epsilon_C$ this amplitude becomes exponentially small and
so these terms can be neglected in the effective Hamiltonian (\ref{H_NA}).

Thus, we conclude that the array shown in Fig. \ref{NonAbelianInsulatorLattice} provides a
realization of the non-Abelian gauge theory discussed in Sec. \ref{DNAGT},
if $\epsilon_J \ll
\epsilon_C \ll E_C \lesssim E_J$ and
\begin{equation}
\frac{\epsilon_J^{2}}{2\epsilon_C} \gtrsim r_0 \gg
\frac{1}{20} \left( \frac{3 \epsilon_J}{4 \epsilon_C} \right)^{5}
\epsilon_J.
\label{PhysCond}
\end{equation}

\subsection{Kinematical Gauss law implementation}
\label{subB}

As we argue below, the straightforward implementation of the
kinematical non-Abelian Gauss law
is not possible because it does not protect the system from the coupling to the
environment. In this respect, the non-Abelian theories are crucially different
from the Abelian ones. In order to explain this difference, we first recollect
the physical arguments that show that Abelian arrays (similar to the ones
suggested in Ref. \cite{Ioffe2002}) are optimally protected from the environment and
then show why non-Abelian ones are different.

For Abelian theories, the Gauss law (\ref{DiscreteGauss}) and (\ref{Gauss_NA}) is even easier
to implement in the physical system than a gauge theory. As discussed
in Sec. II,
in order to implement an Abelian group $\nbZ_n$ on the triangular lattice shown
in Fig.~1 one might assign elements (masus discussed in the
preceding subsection)
that are $2\pi/n$ periodic in superconducting phase to the links of
the dual hexagonal
lattice. In this system the 'site' representation acquires a
transparent physical meaning
-- the state of each island is described by the superconducting phase
of each site of the
hexagonal lattice. Different topological sectors of the system with
hole(s) correspond
to different phases of the inner boundary. The existence of the
protected subspace is
related to the $U(1)$ physical invariance: superconducting phase of
each island (and of
the inner boundary) has no meaning by itself, so only the difference
of the phases of the
neighboring sites might be coupled to physical operators.

The operator conjugated to the superconducting phase is the physical
charge of the whole island and this presents some danger since it can be
coupled to the external electric fields, especially if it results in a direct
coupling of the state of the inner boundary to the electrical fields.
Such a coupling, however, becomes exponentially small in the system with a long
range order in superconducting phase, more precisely in the systems where
$\langle \exp(i n\phi) \rangle$ is close to $1$.
In these systems the total charge of the boundary fluctuates strongly,
while the conjugate operators appearing in the discrete gauge theory
correspond to the
total charge modulo $n$ and thus are decoupled from the physical
electric fields. We compute their weak residual effect on the
degeneracy of the protected
states in Appendix D and find that it quickly becomes exponentially
small [see Eq. (\ref{delta_Q})]
with the number of links $Z$ connecting the boundary to the inside
of the array, i.e.,
in the thermodynamic limit.

\begin{figure}[htb]
\includegraphics[width=3.0in]{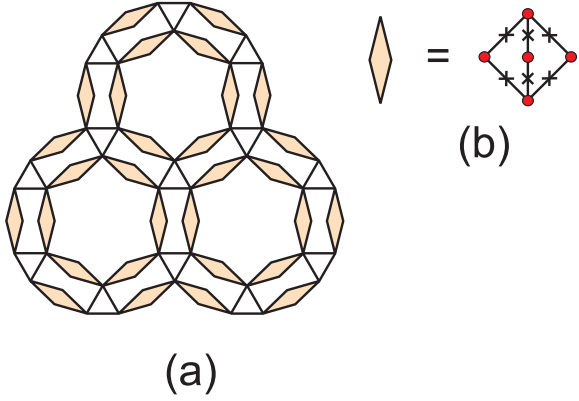}
\caption{
(a) Implementation of the non-Abelian theory with group $D_3$.
The elementary block
contains two elements (called masus and shown as rhombi),
each element is periodic in the phase across it with
periodicity $2\pi/3$, and one triangle which contains
Josephson junctions in each side and is
subjected to magnetic flux $\Phi_0/2$. This ensures that two phases
on the ends of
each block are different. The construction of each masu is shown
separately in (b). The
periodicity is achieved applying magnetic field with flux $\Phi_0/3$
through each
elementary triangle.
}
\label{PhaseArray}
\end{figure}

As mentioned in the preceding section, the most straightforward
way to generalize
this $\nbZ_n$ construction to the simplest
non-Abelian group $S_3$ would be to construct the bonds  characterized by
the elements of the permutation group and the sites that can be
viewed as results
of these permutations. This is achieved in the array shown in Fig.
\ref{PhaseArray}
where bonds are constructed from  pairs of masus and the sites from triangles
frustrated by the magnetic field. This array differs from the one shown in
Fig. \ref{NonAbelianInsulatorLattice} by the absence of the weak links.
All classical states of this array are uniquely specified by the
configurations of
permutation group elements which satisfy the constraint $\prod_{hex.}
g_{ab}=e$.
A small but nonzero charging energy of each contact would lead
to the quantum transitions between different classical states as
described by the Hamiltonian (\ref{H_Gauss_NA}). In this physical
system, these transitions
correspond to the tunneling of the phases on each triangle from one
configuration to another.

%
%
%
%

Thus, in ideal conditions, this system is described by the kinematic
Gauss law model
of preceding section. The situation is not so good for a realistic
system because in
such implementation, one can not exclude the appearance of physical
operators coupled directly
to the  state of each triangle. In particular, this could distinguish
between states with
different chiralities (i.e., in which the phase winds clockwise or
counterclockwise around the
triangle). This is bad news for the effective gauge field model since
the phase differences on distant triangles are related by
%
%
\begin{equation}
\Delta \phi_1 = \Delta \phi_2 + \frac{2 \pi}{3} \text{sign}\left(\prod_C g_{ab}
\right),
\end{equation}
%
%
where $\text{sign}$ denotes the sign of the permutation and indices
$1$ and $2$ denote two distant triangles.
Thus, the terms in the energy that depend on the phase difference
$\Delta \phi$ results in nonlocal terms in the effective Hamiltonian
when expressed as a function of the gauge fields $g_{ab}$. Such terms
ruin the topological
protection, especially if these triangles belong to inner and outer borders.
To be more specific we shall discuss below the effect of
%
%
\begin{equation}
V_{dis.}=V_d \cos(\Delta \phi_1 - \Delta \phi_2)
\mbox{,}
\label{V_dis}
\end{equation}
%
%
which is actually the most dangerous term of this type since it
changes the relative energies of different topological sectors of the
array shown in Fig. \ref{PhaseArray} by an amount proportional to $V_d$.

The effect of $V_{dis.}$ on the array with weak links (discussed in
Sec. \ref{subA})
is exponentially weak because in this system the relation between
phases also contains
a contribution from the continuous phases across weak junctions
which, in the absence
of the superconducting long range order, fluctuate strongly and eliminate the
correlation between $\Delta \phi_1 - \Delta \phi_2$ and $\prod_C g_{ab}$.
Indeed, in this case, the relation between $\Delta \phi_1$ and $\Delta \phi_2$
acquires an additional term which is due the continuous part of the phase
$\tilde{\phi}$ across the weak links.
%
\begin{equation}
\Delta \phi_1 - \Delta \phi_2  = \frac{2 \pi}{3}\text{sign}(\prod_C g_{ab}) +
\sum_C \delta \tilde{\phi}
\mbox{,}
\label{Delta_phi}
\end{equation}
%
where $\delta\tilde{\phi}$ is the difference of continuous phases across two
weak links constituting one bond of the effective lattice. In the insulating
limit of this model ($\epsilon_C\gg\epsilon_J$) the continuous phases fluctuate
practically independently on each bond with
$\langle \cos(\delta \tilde{\phi}) \rangle \sim (\epsilon_J/\epsilon_C)$,
suppressing the effect of the noise (\ref{V_dis})
%
%
\begin{equation}
V^{eff}_{dis.} \sim V_{d} \left( \frac{\epsilon_J}{\epsilon_C} \right)^L
\cos \left[ \frac{2 \pi}{3}\text{sign} \left(\prod_C g_{ab}\right) \right]
\mbox{,}
\label{V_eff}
\end{equation}
%
%
where $L$ is the length of the contour connecting triangles $1$ and $2$.

Alternatively, one can argue that the mapping of the physical system onto
the effective model with variables defined on the bonds of the lattice
does not produce nonlocal (dangerous) operators \textit{if} the states
of the site are indistinguishable. In the insulating array, this is achieved
by making the phase of each triangle (effective site) fluctuate strongly.
In the Abelian $\nbZ_n$ array, this is true because absolute values of the
phase (which distinguish between different states of a given site)
are not measurable,
so generally a system in which each site is characterized only by a single
phase is a good candidate for an array that can be mapped to a bond model.
The problem is that the straightforward implementations of this idea lead
to a $\nbZ_n$ group.

\begin{figure}[htb]
\includegraphics[width=3.0in]{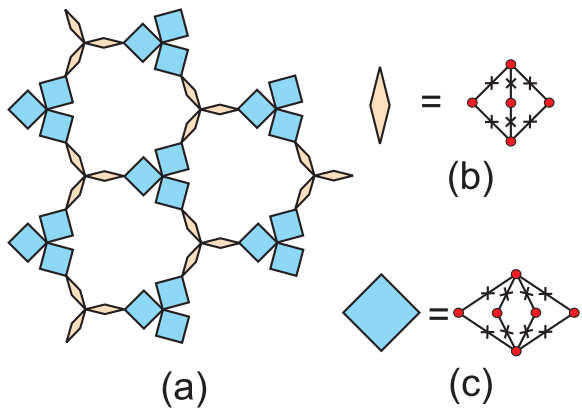}
\caption{
(a) Robust implementation of the non-Abelian group $D_4$
involving elements in phase and charge regime.
(b) Rhombi denote elements that are $2 \pi/3$ periodic and whose charging
energy is relatively high.
(c) Big squares denote elements that are $2 \pi/4$ periodic
in phase, the junctions of these elements are in a quasiclassical regime.
}
\label{PhaseChargeArray}
\end{figure}

In order to add a reflection to the $\nbZ_n$ group, one needs states that
are characterized by both the discrete value of the phase and the
sign of the conjugate
variable, $q$. This is possible to achieve if one starts for the system where
$n \times m$ discrete phases are allowed, and include a large interaction
that forms a degenerate doublet of chiral states from the subset of
$m$ discrete states.
To be more specific, consider the low-energy states that are
described by the wave functions
%
%
\begin{equation}
\Psi_{\pm}(l) = \sum_{k=0}^{k<m} \exp(\pm 2\pi i k / m ) |l,k\rangle, 
\end{equation}
%
%
for $l<n$.
Because the reflection changes the chirality of each state, it is not
reducible to a
simple rotation, so that combined with the phase rotations, it gives
the fundamental
representation of $D_n$ group. We now discuss the construction of the
simplest array that
has these general features. The minimal value of $m$ that allows one
to distinguish
left and right states is $m=3$. The minimal value of $n$ that allows
a non-Abelian group
is
$n=3$, however, $n=m$ is very difficult to implement because it can not
be achieved connecting in series the elements with $n$ and $m$
periodicity (that would
give $n=m$ periodicity instead of $n^{2}$ one). Thus, the simplest
array of this type
corresponds to $n=4$ and group $D_4$. A possible realization is shown in
Fig. \ref{PhaseChargeArray}. Here each link consists of two elements which
have rather different values of $E_J/E_C$: the masu
that is $2 \pi/4$ periodic operates in the phase regime with
$E_{J4}/E_{C4}>1$ while the masu that is $2 \pi/3$ periodic has a
different relation
$E_{J3}/E_{C3} \sim 1$.
More precisely, the amplitude mixing states that differ by $2\pi/4$ is
%
%
\begin{equation}
r \approx E_{J4}^{3/4} E_{C4}^{1/4} \exp(-3 S_4),
\end{equation}
%
%
where $S_4 =\sqrt{E_{J4}/E_{C4}}$ must be much smaller than the energy gap 
%
%
\begin{equation}
\tilde{\Delta} \approx E_{J3}^{3/4} E_{C3}^{1/4} \exp(-3 S_3) 
\end{equation}
%
%
between the doublet of the
chiral states and the remaining $p=0$ state of the $m=3$ set ($S_3 = \sqrt{E_{J3}/E_{C3}}$).

%
%
%
%

It is crucial for this design that the states with $p=\pm 1$ [here $p$ has the
same meaning as in Sec. III, e.g., in Eq. (\ref{p_ab})]
have equal energies which are less than the energy of the state with
$p=0$. This happens if the transition amplitude between the classical
minima is real and negative.
Generally, the amplitude of the transition that increments the superconducting
phase by $2 \pi /3$  contains the phase factor $2 \pi Q/3$ where $Q$
is the average charge of the superconductor whose phase has been changed
(here and below we measure the charge in the units of Cooper pairs).
We require that this phase is $(\pi + 2\pi N)$ where $N$ is arbitrary integer.
This implies that the system should be biased so that the average
charge of the island that experiences this transition is $Q=3/2 + 3N$.

We now discuss the effect of the induced charge on the amplitude of
the transition
between $l=0, \ldots ,3$ discrete states. For simplicity, we shall
assume that the
islands which experience these transitions coincide with the islands
whose phases
change by $2 \pi /3$ and, thus, have the same charge. These
transitions increment
phase by $\pi/2$; so that their amplitude  acquires a phase $(3/2+3N)\pi/2$; in
the notations of Sec. \ref{DNAGT2},  $r_\zeta = \exp(iQ\pi/2) |r|$.
This amplitude has positive real part for $Q =9/2,15/2,\ldots$.
The transition between states with $p= \pm 1$ appears
naturally if the energies of the classical minima with different $k=0,1,2$ are
not exactly equal because, in the momentum representation, this induces
a transition between these states. Such a difference is induced
by the deviations, $\delta \Phi_3=\Phi_3-\Phi_0/3$ of the actual flux
through the elementary rhombus from its ideal value, $\Phi_0/3$.
Thus, in these conditions the low-energy dynamics of such a system is
described by
exactly (\ref{H_i}) with the parameters $r_\zeta$, and $r_{\tau}=E_{J3} \delta
\Phi_3 / \Phi_0$. We emphasize again that the low-energy sector
has to be separated from higher-energy excitations by a large gap
$\tilde{\Delta} \gg r_\zeta,
r_\tau$.

\section{Conclusion}

In this work, we have proposed several possible implementations of lattice
models based on discrete non-Abelian local symmetry groups. The first key
ingredient in all these constructions is the existence of a large but discrete
manifold of equivalent ``classical" states. This manifold appears if the
system is built from heavily frustrated elementary blocks, such as
Josephson-junction loops threaded by a well adjusted magnetic flux. Further,
one needs to select an interaction between these blocks that imposes a set
of local constraints in the low-energy sector. We have found two inequivalent
(but somewhat similar) types of constraints compatible with the non-Abelian
symmetry group. The first type of constraint corresponds to the dynamical
selection of the states which are invariant under the gauge group
transformations.
This is most naturally implemented in insulating arrays where it can be made
equivalent to the selection of states that minimize the
kinetic energy associated with the local tunneling processes between
classically
equivalent states. The second type of constraint emerges more naturally in
superconducting Josephson-junction arrays (or the ones with small
phase fluctuations),
where group elements associated with the bonds can be viewed as
transformations
between the states of the sites of the array. Because the state of each site is
uniquely defined, the product of group elements around a closed loop
is the unit
element. We have called this condition a kinematical Gauss law.
This constraint becomes equivalent to the local gauge-invariance
condition in Abelian theories where a duality transformation maps one onto the
other (Sec. \ref{DAGT})  but for a non-Abelian symmetry it leads to a
different theory in which the roles of charges and fluxons have been reversed
(Sec. \ref{DNAGT}).

In both types of models that we considered, the elementary
excitations within the
constrained subspace are local in space and have a finite-energy gap. We call
these excitations ``fluxons". Generally, these fluxons can be trapped inside a
larger hole prepared in the array leading to a degenerate ground state in the
thermodynamic limit. In a finite system, this degeneracy is lifted by
an exponentially
weak tunneling escape process. It is also robust with respect to all noises
represented by local operators such as local electric fields (generated by
stray charges) or stray magnetic fluxes: the matrix elements of all these
operators are exponentially suppressed. This remarkable property may open
the way to the interesting applications such as storage of quantum information
for which a low decoherence rate is a must. We emphasize that the physical
nature of these fluxons depends, however, on the details of implementation and
the type of the constraint: the canonical gauge invariance is more easily
implemented in insulating arrays in which fluxons have the physical
meaning of fractional vortices while kinematical Gauss law is more naturally
implemented in superconducting arrays in which fluxons carry physical charges.

Besides fluxons, in any physical array, it is possible to build excitations
that violate a local constraint (of course, at some energy cost); in the
gauge theory language these excitations are pseudocharges. In the canonical
gauge models pseudocharges may provide a multidimensional representation
of the discrete gauge group while fluxons are labeled by their
conjugacy classes.
In the kinematical Gauss law models the role of fluxons and pseudocharges is
reversed. The existence of these higher-dimensional representations leads to
a very interesting statistical interaction between fluxons and pseudocharges
which may be utilized for making exact transformations needed in quantum
computations.

Many interesting questions remain to be addressed, we shall list just a few
here. First, it is not clear whether it is absolutely unavoidable to emulate
gauge theory in order to get the topological protection and non-Abelian
statistics. What is the most general class of theories that has these
properties? Second, is it possible that some simpler overfrustrated arrays
``spontaneously" develop the gauge invariance in the low-energy sector?
\cite{foot3}.
Such an array would be more experimentally accessible and easier to
control. Coming
back to the proposed arrays it would be very interesting to investigate the
experimental manifestations of the nonlocal nature of the fluxon quantum
numbers which become apparent only when a few of them are present. As
discussed extensively in the mathematical literature
\cite{Dijkgraaf1990,Bais1992},
these quantum numbers interpolate between a set of conjugacy classes
(one for each
fluxon) and a set of group elements (which is not a gauge-invariant object
by itself).
It would be important to realize how these subtle but
beautiful concepts can be revealed in a physical experiment, and to
design some setup that could verify the non-Abelian statistics of
the excitations in these
models.
Finally, it would be very interesting to adapt the construction of
the Chern-Simons
term (which binds fluxons and charges) known for a continuous groups
to the discrete finite gauge groups (even Abelian ones) discussed here and to provide a
corresponding physical
implementation.

\begin{acknowledgments}
L.I. is grateful to ENS and LPTHE for their hospitality which allowed this
work to be commenced and finished. This work was partially supported by NSF
Grant No. DMR 0210575.
\end{acknowledgments}

\appendix

\section{Tunneling in a single rhombus}

Here, we calculate the transition amplitude between the
degenerate classical states appearing in a single rhombus with diagonal
that is placed in the magnetic field with the flux $2 \Phi_0 /3$
through the whole rhombus. In the following we shall refer to this object
as masu. As a function of the phase difference $\psi$ across the
diagonal the total Josephson energy of such masu is:
\begin{eqnarray}
E(\psi)&=& - 2 E_J \left[
\left|\cos \left({\psi}/{2}\right)\right|+
\left|\cos\left({\psi}/{2}+{\pi}/{3}\right)\right| \right. \nonumber \\
&&+
\left. \left|\cos\left({\psi}/{2}-{\pi}/{3}\right)\right|
\right].
\end{eqnarray}
It is periodic in $\psi$ with the period $2\pi/3$ and has three classically
degenerate minima: $\psi=0,\pm 2\pi/3$.

\begin{figure}[ht]
\includegraphics[width=2.6in]{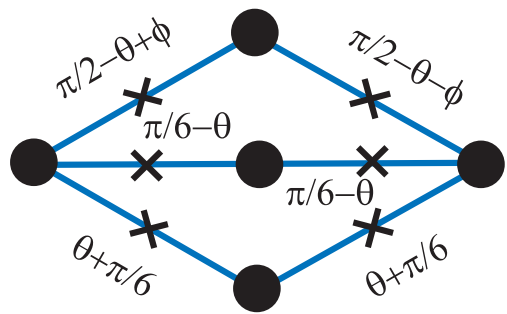}
\caption{
Tunneling in the elementary building block of non-Abelian arrays.
In the quasiclassical approximation, tunneling is described by the
simultaneous evolution of all phases that are parametrized by
two independent variables, $\theta$ and $\phi$.
}
\label{Masu}
\end{figure}

The amplitude of transitions $r_m$
between these minima can be computed in a
quasiclassical approximation, as $r_m \approx E_J^{3/4} E_C^{1/4} \exp(-S_0)$,
and is equal for all transitions. To calculate the action $S_0$, we choose two
independent superconducting phases as shown in Fig. \ref{Masu} and rescale
$t$ into $t/ \sqrt{E_J E_C}$. In terms of these variables the action is
\begin{eqnarray}
S &=& \int \left[ \frac{3}{8} \left( \frac{d\theta}{dt} \right)^2 +
     \frac{1}{8} \left( \frac{d\phi}{dt} \right)^2 -
     2 \sqrt{3} \cos (\theta ) \right. \nonumber \\
&&
 - 2 \cos (\phi )\,\sin (\theta ) \Bigg] dt
\mbox{.}
\end{eqnarray}
The minimum of this action corresponds to the quasiclassical equations
of motion:
\begin{eqnarray}
\frac{d^2 \theta}{dt^2} &=&
\frac{8}{\sqrt{3}} \sin (\theta ) - \frac{8}{3} \cos (\theta ) \cos (\phi),  \\
\frac{d^2 \phi}{dt^2} &=& 8 \sin (\theta ) \sin (\phi ).
\end{eqnarray}

Solving these equations numerically, we get the expression for the transition
amplitude in one masu
\begin{equation}
r_{m} \approx E_J^{3/4} E_C^{1/4} \exp(-S_0) \; \; \;\; S_0= 1.22
\sqrt{\frac{E_J}{E_C}}.
\label{r_m}
\end{equation}

%
%
%
%

\section{Tunneling of four rhombi and triangle}

\begin{figure}[ht]
\includegraphics[width=3.2in]{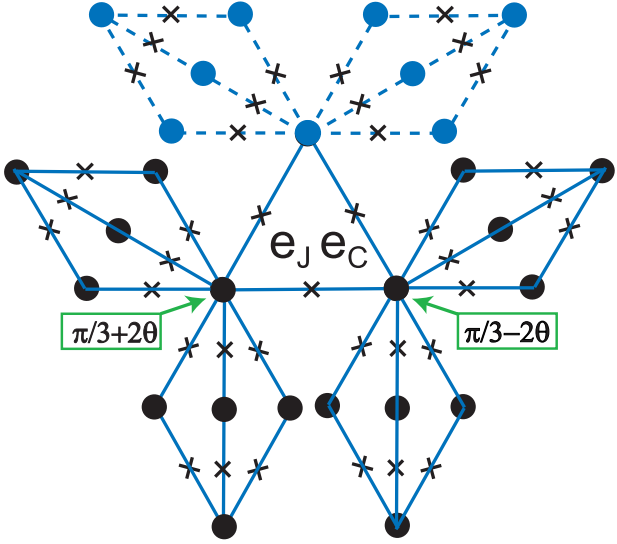}
\caption{
Basic tunneling process in the full lattice.
}
\label{FourMasu}
\end{figure}

\begin{figure}[ht]
\includegraphics[width=3.2in]{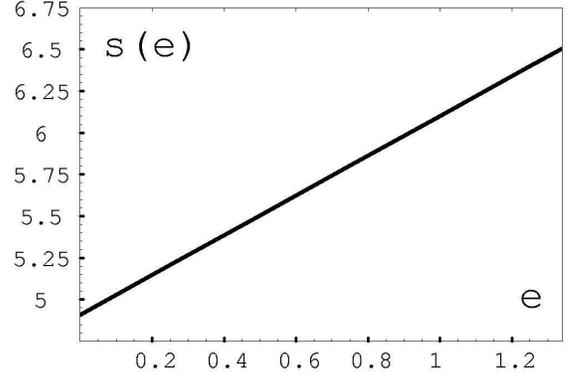}
\caption{
Dimensionless action of the tunneling process.
}
\label{s(e)}
\end{figure}

Here, we calculate the transition amplitude of four masus connected by
a triangle as
shown in Fig. \ref{FourMasu}. If the junctions constituting the
triangle are weak,
$e_J\ll E_J$ this amplitude can be obtained in the perturbation
theory as a product
of two transitions that happen consecutively: one involving the left
pair of masus
and another the right one. The amplitude for each transition is related
to the amplitude of the transition in a single masu (\ref{r_m}) because the
quasiclassical equations of motion for the variables describing two
masus joined
at the vertex are the same as for a single masu. Thus, this
tunneling amplitude
contains twice the action (\ref{r_m}):
\begin{equation}
r_m' \approx E_J^{3/4} E_C^{1/4} \exp(-S_0) \; \; \; \; S_0=2.44
\sqrt{\frac{E_J}{E_C}}
\mbox{.}
\label{r_m'}
\end{equation}
In this limit, the amplitude of the full process is
\begin{equation}
r_0 \approx  \frac{(r_m')^2}{e_J}
\mbox{,}
\label{r_t}
\end{equation}
where again $e_J$ denote the weak Josephson couplings inside the triangles.
%
%
In a general case of triangle junctions comparable with the junctions
constituting
masus, we need to add these terms to the action and solve the full
set of equations of
motion in order to obtain the action corresponding to this process.
The solution of these equations describes a simultaneous variation
of phases in all four masus, so the most general action has the form
\begin{eqnarray}
S&=&4 \sqrt{\frac{E_J}{E_C}} \int \Bigg[ \frac{3}{8} \left(
\frac{d\theta}{dt} \right) ^2 +
     \frac{1}{8} \left( \frac{{d\phi}}{{dt}} \right)^2 - 2 {\sqrt{3}}
\cos (\theta ) \nonumber \\
&& -
     2 \cos (\phi ) \sin (\theta ) + \frac{3 E_C}{8 e_C}
     \left( \frac{d\theta }{dt} \right)^2 \nonumber \\
&&+
     \frac{e_J}{4E_J} \left[ -2 \cos (2 \theta ) + \cos (4 \theta )
\right]  \Bigg].
\end{eqnarray}
In the most natural case the junctions of the triangle have the same
Josephson current per area and
the same capacitance per area as the junctions of the masu, so
$e_J/E_J=E_C/e_C\equiv e$. Minimizing the
action, we get the equations of motion
\begin{eqnarray}
\frac{d^2 \theta}{dt^2} &=& \frac{4}{1+e} \Bigg[
\frac{2}{\sqrt{3}} \sin (\theta ) - \frac{2}{3} \cos (\theta ) \cos (\phi) \nonumber \\
&&
- \frac{e}{3} \left[ \sin(4 \theta) - \sin(2 \theta) \right] \Bigg],  \\
\frac{d^2 \phi}{dt^2} 
&=& 8 \sin (\theta ) \sin (\phi ).
\end{eqnarray}
Solving these equations numerically, we obtain the dimensionless
action $S(e)$ as a function of $e$
shown in Fig. \ref{s(e)} which should be used in
\begin{equation}
r_0 \approx E_J^{3/4} E_C^{1/4} \exp\left(-S(e) \sqrt{\frac{E_J}{E_C}}\right)
\mbox{,}
\label{r_r}
\end{equation}
to obtain the value of the transition amplitude.
We observe that the action $S(e)$ varies only by 25\% [from $S(0)=4.91$ to
$S(1)=6.10$] when $e$ is increased
from $0$ to $1$, so the amplitude of the transition is not too
sensitive to the strength of Josephson
junctions in the triangles.

%
%
%

\section{Effective Josephson coupling of a chain}

Here we calculate the effective coupling by a chain of $(n+1)$ weak Josephson
contacts. Denoting the phase across each contact by $\varphi _{i}$,
the phase across
the last contact is $\phi -\sum_{i}\varphi _{i}$. After a time rescaling
the system is described by a dimensionless Lagrangian
\begin{eqnarray}
L &=&L_{0}+L_{J}, \\
L_{0} &=&\frac{1}{2}\sum_{i}\left( \frac{d\varphi _{i}}{dt}\right) ^{2}+%
\frac{1}{2}\left( \sum_{i}\frac{d\varphi _{i}}{dt}\right) ^{2}, \\
L_{J} &=&\lambda \left[ \sum_{i}\cos \varphi _{i}+\cos \left( \phi
-\sum_{i}\varphi _{i}\right) \right],
\end{eqnarray}
where $\lambda =\epsilon_{J}/(8\epsilon_{C})$.
Let us assume that $\lambda \ll 1$ and
treat Josephson energy in perturbation theory. The bare Hamiltonian is
quadratic in charges, $q_i$, conjugated to the phases $\varphi _{i}$
\begin{equation}
H_{0}=\frac{1}{2}\left[ \sum_{i}q_{i}^{2}-\frac{1}{n+1}\left(
\sum_{i}q_{i}\right) ^{2}\right].
\end{equation}
The ground state of this Hamiltonian corresponds to all charges $q_{i}=0$
while in the low-energy states some $q_{i}=\pm 1$. The energy of the state
with $k$ charges $q_{i}=1$ is 
\begin{equation}
\epsilon _{k}=k(n+1-k)/[2(n+1)].
\end{equation}
To find the effective coupling provided by this chain we need to calculate  
\begin{equation}
W(\phi) = -\lambda \left\langle \cos \left( \phi -\sum_{i}\varphi
_{i}\right) \right\rangle,
\end{equation}
in the lowest order of the perturbation theory in
\begin{equation}
V_{int}=\frac{\lambda}{2} \sum_i  \exp(i \varphi_i) + h.c.
\end{equation}
In the charge representation, one has
\begin{equation}
W(\phi)=-\lambda\left\langle \prod_{i}q^{+}_{i}\right\rangle \cos \phi
\end{equation}
where $q^{+}_{i}$ is the operator increasing the charge of junction $i$ by $1$ and
\begin{equation}
V_{int}=\frac{\lambda}{2} \sum_i q^{+}_i + h.c.
\end{equation}
In the leading order, we can limit ourselves to the components of the ground-state wave functions that
contain charges $-1$ at some $k$ junctions and $(n-k)$ opposite charges at the remaining junctions.
These components have coefficients  $\Psi _{k}$ that can be found from a
recursive relation
\begin{equation}
\Psi _{k}= \frac{k \lambda}{2 \epsilon_k} \Psi _{k-1},
\end{equation}
so we get 
\begin{equation}
\Psi _{k}=\frac{(n-k)!}{n!} (n+1)^k \lambda ^{k}.
\end{equation}

Expressing the average $\left\langle \prod_{i} q^{+}_{i} \right\rangle$ through these components via
\begin{equation}
\left\langle \prod_{i} q^{+}_{i} \right\rangle=
\sum_{k=0}^n \frac{n!}{(n-k)! k!} \Psi _{k} \Psi _{n-k}
\mbox{,}
\end{equation}
we get
\begin{equation}
\left\langle \prod_{i}q^{+}_{i}\right\rangle =\frac{(n+1)^{n+1}}{n!}
\lambda ^{n}.
\end{equation}
Converting back to unrescaled variables we conclude that the effective coupling
between the ends of the line consisting of $n$ junctions is
\begin{equation}
W_n(\phi) = -\frac{n^n}{(n-1)!}
\left( \frac{\epsilon_{J}}{8\epsilon_{C}}\right) ^{n-1}\epsilon_{J}\cos \phi.
\label{W_n}
\end{equation}

\section{Effect of the electric noise on the protected subspace in
$\nbZ_n$ arrays}

Quantitatively, consider the island coupled to $Z$ links by $2\pi/n$
periodic elements.
Each element is characterized by the potential energy
$V_{el}(\phi)=\sum_k V_{0}(\phi+2\pi k/n)$
which has minima at $\phi_m=2\pi (m+\frac{1}{2})/n$.
Near each minima the potential is approximately Gaussian:
\begin{equation}
V_{el} \approx \frac{c}{2} E_J (\phi-\phi_m)^2,
\end{equation}
where
\begin{equation}
c= \frac{1}{2}\sum_{k=0}^{n-1} \Bigg|\cos\left[\frac{\pi}{n} \left( k+\frac{1}{2} \right)\right] \Bigg|,
\end{equation}
is a numerical coefficient that
is equal to $1/\sqrt{2}$ for $n=2$ and $n/\pi$ for large $n$. The
dynamics of small phase
fluctuations is described by the Lagrangian
%
%
\begin{equation}
\mathcal{L}=\sum_{(ij)} \left( \frac{n}{32 E_C}(\dot{\phi_i} -
\dot{\phi_j})^2 + V_{el}(\phi_i - \phi_j) \right)
\mbox{,}
\end{equation}
%
%
where $E_C=e^2/(2C)$ is the charging energy of each junction.
In this approximation, each phase difference
is an independent Gaussian variable
which is described by the wave functions
%
%
\begin{equation}
\Psi(\delta \phi)=\exp(-\frac{1}{2}\sqrt{\frac{nc E_J}{16E_C}}\delta
\phi^2 ).
\end{equation}
%
%
The canonically conjugate charge fluctuations are therefore also
Gaussian with width
$\sqrt{(nc E_J)/(16E_C)}/2$;
charge fluctuations on different junctions add up to the total charge
fluctuation on the island with the Gaussian width
%
%
\begin{equation}
\langle (Q-\bar{Q})^2 \rangle = \frac{Z}{8} \sqrt{\frac{nc E_J}{E_C}}.
\end{equation}
%
%
If these fluctuations are restricted to a given value modulo $n$, they give
slightly different results.
Consider, for instance, the difference between average square of the charge
between the sectors with different $Q$ modulo $n$
%
\begin{equation}
\langle \delta Q^2 \rangle_k - \langle \delta Q^2 \rangle_{l} \sim
\exp\left(-\frac{\pi^2 Z}{4}
\sqrt{\frac{c E_J}{n^3 E_C}}\right).
\label{delta_Q}
\end{equation}
%
This difference quickly becomes exponentially small for a large number
of links connecting a given island. Because a topologically protected
space can be viewed as spanned by
different states of the inner boundary of the array connected to the
interior by $Z \gg 1$ links, the electrical fields have exponentially
weak effects on the topologically protected subspace.


\begin{thebibliography}{99}

\bibliographystyle{prsty}

\bibitem{Ekert1996} A. Ekert and R. Jozsa, Rev. Mod. Phys.
\textbf{68}, 733 (1996).

\bibitem{Steane1998} A. Steane, Rep. Prog. Phys. \textbf{61}, 117 (1998).

\bibitem{Preskill1998} J. Preskill, Proc. R. Soc. London A \textbf{454}, 385 (1998).

\bibitem{Wen1990}  X. G. Wen and Q. Niu, Phys. Rev. B \textbf{41}, 9377 (1990).

\bibitem{Wen1991}  X. G. Wen, Phys. Rev. B \textbf{44}, 2664 (1991).

\bibitem{Read1991} N. Read and S. Sachdev, Phys. Rev. Lett. \textbf{66},
1773 (1991).

\bibitem{Ioffe2002} L. B. Ioffe and M. V. Feigel'man, Phys. Rev. B
\textbf{66}, 224503 (2002).

\bibitem{Doucot2003} B. Dou\c{c}ot, M. V. Feigel'man, and L. B. Ioffe,
 Phys. Rev. Lett. {\bf 90}, 107003 (2003).

\bibitem{Kitaev1997} A. Kitaev, Annals of Physics {\bf 303},  2  (2003).

\bibitem{Propitius1999} M. de Wild Propitius and F. A. Bais,
in \textit{Particle
and  Fields}, edited by G. W. Semenoff and L. Vinet, CRM series in
Mathematics, (Springer-Verlag, New York 1999),  p. 353.

\bibitem{Odintsov}
A.~A. Odintsov and Yu.~V. Nazarov, Phys. Rev. B {\bf 51},  1133  (1995).

\bibitem{Choi}
M.~Y. Choi, Phys. Rev. B {\bf 50},  10088  (1994).

\bibitem{Stern}
A. Stern, Phys. Rev. B {\bf 50},  10092  (1994).

\bibitem{Diamantini}
M.~C. Diamantini, P. Sodano, and C.~A. Trugenberger, Nucl. Phys. B {\bf 474},
  641  (1996).

\bibitem{Doucot2002}  B. Dou\c{c}ot and J. Vidal, Phys. Rev. Lett.,
\textbf{88}, 227005 (2002).

\bibitem{Mermin1979} N.~D. Mermin, Rev. Mod. Phys. \textbf{51}, 591 (1979).

\bibitem{Bais1980} F. A. Bais, Nucl. Phys. \textbf{B 170}, 32 (1980).

\bibitem{Dijkgraaf1990} R. Dijkgraaf, V. Pasquier, and P. Roche, Nucl. Phys. B,
Proc. Suppl. \textbf{18}B, 60 (1990).

\bibitem{Bais1992} F. A. Bais, P. van Driel, and M. de Wild
Propitius, Phys. Lett. B
\textbf{280}, 63 (1992).

\bibitem{Mochon1}
C. Mochon, Phys. Rev. A {\bf 67},  022315  (2003).

\bibitem{Mochon2}
C. Mochon, Phys. Rev. A {\bf 69},  032306  (2004).

\bibitem{foot1} Generally, the operator $F(g_{ab})$ does not
necessarily conserve the total number of fluxons. But in the limit
where this term is small compared to the first term of $H$, we may, to
lowest nontrivial order in perturbation theory, project it onto the
subspace of one-fluxon states, i.e. to replace it by $\hat t(g_{ab})$.

\bibitem{foot2} In Japanese ``masu" is used to describe the shape of rhombus
with diagonal and
patterns build from it, we are thankful to H. Kojima for suggesting
this term to us.

\bibitem{foot3} A somewhat similar phenomenon does occur in classical wire arrays,
(see \cite{Park2001}).

\bibitem{Park2001} K. Park and D. A. Huse, Phys. Rev. B \textbf{64},
134522 (2001).



\end{thebibliography}

\end{document}